\documentclass[12pt,a4paper,final]{iopart}

\usepackage{iopams}  
\usepackage{graphicx}
\usepackage[breaklinks=true,colorlinks=true,linkcolor=blue,urlcolor=blue,citecolor=blue]{hyperref}

\usepackage{bm}

\begin{document}
\title[Uncertainty Quantification in Lattice QCD Calculations for Nuclear Physics ]{Uncertainty Quantification in Lattice QCD Calculations for Nuclear Physics}

\author{Silas R. Beane}
\address{Department of Physics, University of
  Washington, Box 351560, Seattle, WA 98195, USA}
\ead{silas@uw.edu}

\author{William Detmold}
\address{Center for Theoretical Physics, Massachusetts Institute of Technology, Cambridge, MA 02139, USA}
\eads{wdetmold@mit.edu}

\author{Kostas Orginos$^{1,2}$}
\address{$^1$Department of Physics, College of
  William and Mary, Williamsburg, VA 23187-8795, USA}
\address{$^2$Jefferson Laboratory, 12000 Jefferson Avenue, Newport
  News, VA 23606, USA}
\ead{kostas@jlab.org}

\author[cor1]{Martin J. Savage}
\address{Institute for Nuclear Theory, Box 351550, Seattle, WA 98195-1550, USA}
\eads{mjs5@uw.edu}

\begin{abstract}
  The numerical technique of Lattice QCD holds the promise of
  connecting the nuclear forces, nuclei, the spectrum and structure of
  hadrons, and the properties of matter under extreme conditions with
  the underlying theory of the strong interactions, quantum
  chromodynamics.  A
  distinguishing, and thus far unique, feature of this formulation is
  that all of the associated uncertainties, both statistical and
  systematic can, in principle, be systematically reduced to any desired precision
  with sufficient computational and human resources.
  We review the sources of uncertainty inherent in
  Lattice QCD calculations for nuclear physics, and discuss
  how each is quantified in current efforts.
\end{abstract}

\vspace{2pc}

\section{Introduction}
\noindent
Quantum chromodynamics (QCD) and the electroweak interactions are
responsible for the nuclear forces, and consequently for the structure
and interactions of all nuclei.  
Historically, the complexity of  QCD  has prevented  direct calculation of
the properties of
low-energy and medium-energy nuclear systems.
However, after decades of development, Lattice QCD (LQCD), a technique to
numerically evaluate the QCD path integral, promises to permit QCD
calculations of low-energy nuclear processes with uncertainties that
can be systematically reduced to any desired precision with sufficient
human and computational resources.  Before having complete confidence in 
LQCD predictions for nuclear physics, it is critical to verify it as a
reliable technique.  
This will be accomplished by demonstrating
agreement with a diverse array of experimental measurements, 
and showing that the uncertainties of the LQCD
calculations behave as expected with, for instance, increasing lattice
volumes, numbers of gauge-field
configurations and decreasing lattice spacings.

As with any meaningful  prediction, the uncertainties
associated with a LQCD calculation define its utility.
A complete quantification of all of the uncertainties
associated with any given calculation is essential for it to be 
scientifically complete and provide more than a calculational
benchmark.  During the last few years, the LQCD community has
self-organized and assembled a compendium of lattice results,
mainly of
importance for particle physics~\cite{Aoki:2013ldr}, including a
comprehensive analysis of all associated uncertainties.  
This compendium 
represents the consensus of the entire community and 
is along the same lines as the Particle Data Group
summary
of experimental results in high-energy physics.
It is a valuable
resource,  both within and outside of the LQCD community.
Few quantities of  importance to nuclear physics currently appear in this compendium as 
many calculations 
remain in exploratory stages. 
As these calculations mature, we expect they too will be added to the LQCD compendium.

In this article, we outline the array of techniques required to
perform LQCD calculations for Nuclear Physics, and identify the
uncertainties that are inherent in each of these techniques.  We discuss
the procedures and uncertainties associated with generating the
configurations of gluons fields, with the generation of the
correlations between quarks and gluons, and finally with the
extraction of physical information from hadronic and nuclear correlations.
Our presentation is aimed at nuclear physicists who are not experts in lattice field theory methods.

\section{Lattice QCD Technology}

\subsection{The QCD Path Integral}

\begin{figure}[htb!]
 \centering
 \includegraphics[width=0.55\textwidth]{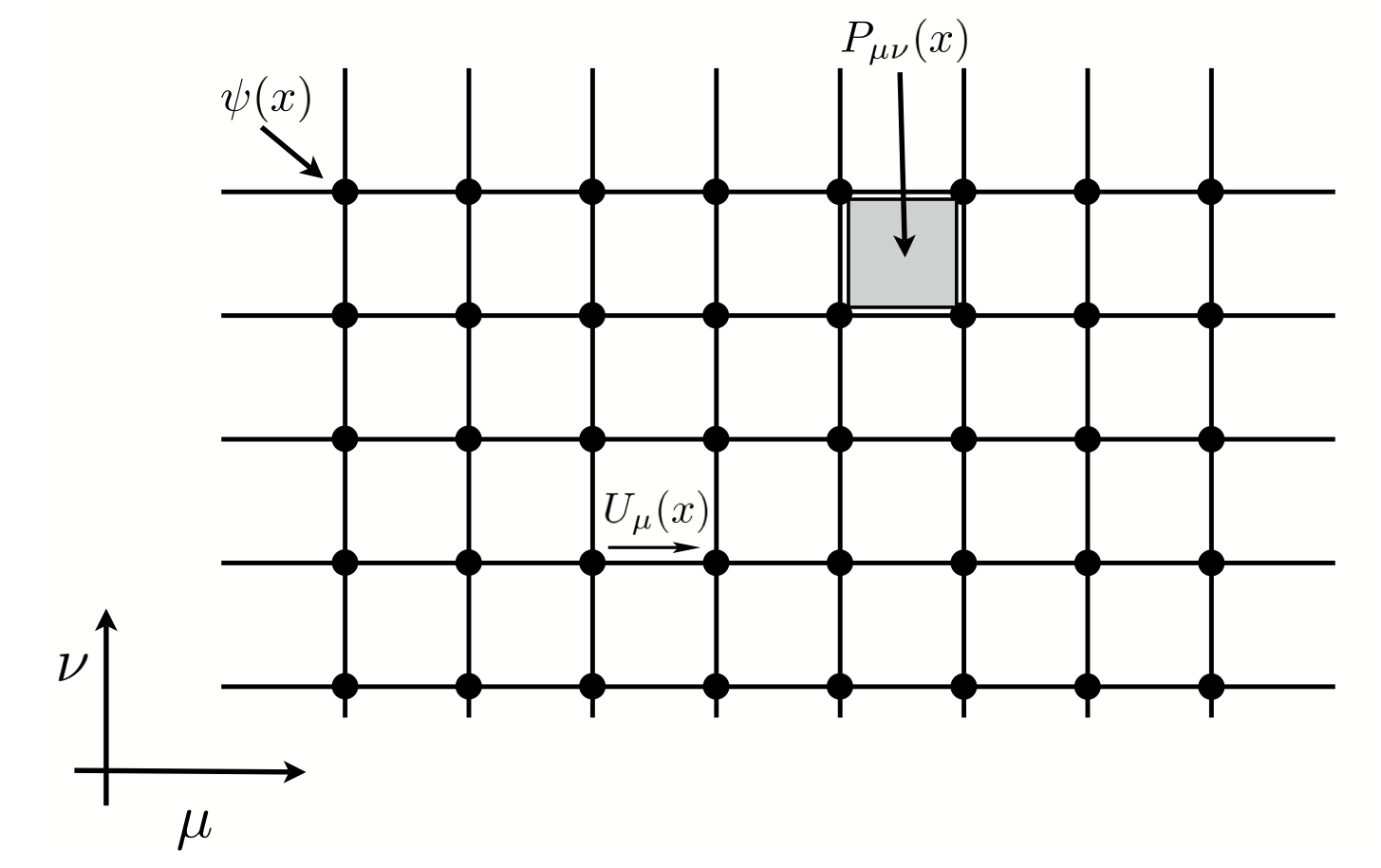}
 \caption{
 A two dimensional slice of the four dimensional  spacetime lattice.
$\hat\mu$ and $\hat\nu$ denote unit vectors in the indicated directions.
$\psi(x)$ denotes a quark field at the lattice site $x$, 
$U_\mu(x)$ denotes the gauge link from the lattice-site $x$ to the site
$x+a\hat\mu$, and $P_{\mu\nu}(x)$ denotes the $1\times 1$ Wilson plaquette centered
at $x+a\hat{\bf \mu}/2+a\hat{\bf \nu}/2$. }
 \label{fig:lattice}
\end{figure}

\noindent
There is only one known way to rigorously define QCD non-perturbatively, and that is as the
continuum limit of a lattice gauge theory. The spacetime lattice provides 
an ultraviolet regulator of the continuum field theory and admits
numerical evaluation of the functional integrals required for
calculating physical observables. 
The QCD partition function is
\begin{equation}
 {\cal Z} = \int {\cal D} A_\mu {\cal D} \bar \psi {\cal D}  \psi \; 
e^{\int d^4 x \left(-\frac{1}{4} G^a_{\mu\nu} G^{a\mu\nu} - 
\sum\limits_f \overline{\psi}_f \left[ D_\mu \gamma_\mu + m_f\right] \psi_f\ 
\right)}
\ \ ,
\end{equation}
where $A_\mu$ is the QCD gauge field (describing the gluons),
$G^a_{\mu\nu}$ is the gauge field strength and $\overline{\psi}_f$, $\psi_f$ are
the quark fields representing  quarks of  flavor $f$.  $D_\mu$ is the QCD covariant
derivative and $\gamma_\mu$ are
the Dirac matrices.  
Physical observables are calculated from correlation
functions of operators, $\cal O$, that are functions of the quantum
fields (quarks and gluons), 
generically written as
\begin{equation}
\langle {\cal O}\rangle = \frac{1}{\cal Z} \int {\cal D} A_\mu {\cal D} \bar
\psi {\cal D}  \psi \; {\cal O}\; e^{\int d^4 x \left(-\frac{1}{4} G^a_{\mu\nu}
    G^{a \mu\nu} - 
\sum\limits_f  \overline{\psi}_f \left[  D_\mu \gamma_\mu + m_f\right] \psi_f\ 
 \right)} \ .
 \end{equation}
The functional  integrals above  
require regularization and 
can be straightforwardly defined on a 
discrete spacetime
that we will take to be a regular hypercubic lattice.
In order to preserve gauge invariance, the gauge
fields are discretized as $SU(3)$ matrices,
$U_\mu (x)$,
associated with the links of the lattice (see
Figure~\ref{fig:lattice}).
The simplest discretized 
form of the 
gauge action is
the sum over all plaquettes,
 $P_{\mu\nu}(x)$,
formed from the product of links, 
$U_\mu(x) = \exp\left( i \int_x^{x+\hat\mu}\ dx^\prime  A_\mu (x^\prime)\right)$,
around  elementary plaquettes of the
lattice,
\begin{equation}
S_g(U) ={\beta} \sum_{x\mu\nu}\left( 1 - \frac{1}{3}
{\rm Re }\ {\rm  Tr}\ {P_{\mu\nu}(x)} \right)  
\ \ ,
 \label{eq:lattice_gauge_action}
\end{equation}
with
 \begin{equation}
 P_{\mu\nu} = U_\mu(x)U_\nu(x+\hat\mu)U^\dagger_\mu(x+\hat\nu)U^\dagger_\nu(x)
\,,
 \label{eq:plaquett}
 \end{equation}
 and $\beta$ is the lattice gauge coupling.  
 Taking the naive continuum limit, 
  this action reduces to the familiar continuum gauge
 action, $\ -\int d^4 x \frac{1}{4}\left( G^a_{\mu\nu}(x)\right)^2 $.
 The action in Eq.~(\ref{eq:lattice_gauge_action}) is the
 Wilson gauge action~\cite{Wilson:1974sk}, and while this
 discretization is not unique, it is the simplest.  
 It can be modified
 by adding larger loops of links
 with coefficients appropriately chosen to
 achieve a more rapid approach to the continuum~\cite{Symanzik:1983dc}, which is an essential
 goal of the calculation.

\subsection{Including Quarks}
 
 Including the dynamics of quarks, which are defined on the vertices of the
 lattice, is a challenge.  A naive discretization of the
 continuum action describing a single fermion introduces 16
 lattice fermion flavors in  four dimensions.  
 The additional 15 light lattice fermion degrees of freedom, 
referred to as ``doublers'',
 can be
 avoided through the use of several ingenious formulations of lattice fermions.
 Wilson fermions, which were 
 the first to be
 introduced~\cite{Wilson:1974sk},
 eliminate the doublers by adding an irrelevant dimension five
 operator to the action that lifts the masses of the doublers, leaving
 only one light fermion in the spectrum.  
 However, this approach explicitly
 breaks chiral symmetry and introduces lattice artifacts that scale as
 $O(a)$, where $a$ is the lattice spacing. Kogut-Susskind fermions~\cite{Kogut:1974ag} (staggered
 fermions) provide another way to remove some of the doublers and
 re-interpret the remaining four as four degenerate flavors.  
 In this approach, 
 a $U(1)$ chiral symmetry  remains unbroken and lattice
 artifacts scale as $O(a^2)$.
 Kogut-Susskind fermions become problematic when the required number
 of flavors is not a multiple of four (as is the case for QCD in nature).  One
 approach to deal with this is to take the fourth root of the
 corresponding determinant (see below).  
 Although this ``rooting''
is not  justified at
 non-zero lattice spacing, current numerical evidence suggests that
 the effects are negligible for many quantities.  
 Finally, the domain-wall~\cite{Kaplan:1992bt,Shamir:1993zy,Furman:1994ky} and
 overlap~\cite{Narayanan:1994gw,Neuberger:1997fp} discretizations
 preserve a lattice chiral symmetry at finite lattice spacing and are
 doubler free. Unfortunately, such formulations are significantly more
 computationally expensive.  In all cases, the lattice fermion action
 is of the form, $S_f = \bar \psi D(U) \psi$, where $\psi$ is the
 fermion ``vector'' and $D(U)$ is a sparse 
 matrix~\footnote{In certain
   cases, such as with overlap fermions, the matrix is not sparse but
   has sparse-like properties.}  
 acting on the fermion vector, that
 depends on the gauge field, $U$.

In the case of two quark flavors, the discretized partition function  is
\begin{eqnarray}
{\cal Z} &=& \int \prod_{\mu,x}dU_\mu(x)\prod_x 
d\bar{\psi}d\psi
\;\; e^{-S_g(U)-S_f(\bar{\psi},\psi,U)} 
\nonumber\\
&=&  \int \prod_{\mu,x}dU_\mu(x)\;\;
 {\rm det}\left(D(U)^\dagger D(U)\right)  \;\; e^{-S_g(U)}
 \ \ ,
 \label{eq:contZ}
 \end{eqnarray}
where 
the integrations over the quark fields (represented by Grassmann numbers)
have been performed exactly, resulting in the determinant factor.
Although the quark matrix
$D(U)$ represents one flavor, the determinant ${\rm det}\left(D(U)^\dagger
D(U)\right) $ represents two flavors
as ${\rm det} D(U)^\dagger = {\rm det} D(U)$. 
In the case of correlation functions defining observables, 
integrating over
the quarks gives
\begin{equation}
\langle{\cal O}\rangle = \frac{1}{\cal Z}
 \int \prod_{\mu,x}dU_\mu(x)\;\; {\cal O}(\frac{1}{D(U)},U)\;\;
 {\rm det}\left(D(U)^\dagger D(U)\right)  \;\; e^{-S_g(U)}\, ,
 \label{eq:CorFunc}
 \end{equation}
where the operators  ${\cal O}$  depend on the inverse of the
quark matrix and (possibly explicitly) on the gauge fields.  
The above expressions are only valid in the case of two
flavors of quarks (the up and the down),
assumed to have the same mass for these discussions,
which is a good approximation to the low energy physics of QCD. 
The strange quark is easily accommodated  by including 
${\rm det}\left(D_s(U)\right) = {\rm det}\left(D_s(U)^\dagger D_s(U)\right)^{1/2}$ 
in the partition function,
as are the charm quarks through a similar factor 
(although their effects in quantities of importance to low-energy nuclear physics are expected  to be small).

\subsection{Monte Carlo Evaluation of the Path Integral}

The evaluation of $\langle{\cal O}\rangle$  in  Eq.~(\ref{eq:CorFunc}) is the main numerical task
faced in LQCD calculations. 
The integrations over the gauge fields 
are of extremely large dimensionality,
and are made practical by restricting spacetime to a compact region. 
Given that QCD has a fundamental length 
scale of $\sim1~{\rm fm}$ ($10^{-13}~{\rm cm}$), 
calculations must be performed in lattice volumes 
(volumes are denoted by $V=L^3\times T$, 
where $L$ is the number of lattice sites in each spatial direction and $T$ the number in the temporal direction)
that have a 
physical size $a L \gg 1~{\rm fm}$
in order to control finite volume effects, and with 
lattice spacings  $a \ll 1~{\rm fm}$ in order to be close to the
continuum limit.  
With moderate choices for the volume and  lattice
spacing, a lattice  of $ 64^3\times 256$ sites is currently practical.  
Accounting for the
color and spin degrees of freedom, such calculations involve
$\approx 10^{10}$ degrees of freedom. The only practical way for 
this type of computation to
be done is by  Monte Carlo integration. Fortunately, the
combination of the quark determinant and  gauge action,
\begin{equation}
 {\cal P}(U) =\frac{1}{\cal Z}\  {\rm det}\left(D(U)^\dagger D(U)\right)  
\;\; e^{-S_g(U)}\,,
 \end{equation}
is a positive-definite quantity that can be interpreted as a
probability measure and hence importance sampling 
methods can be used in performing the integrations.
As will be discussed in detail in the next section, 
the basic
procedure is to generate $N$ gauge field configurations $\{U_i\}$ 
representative of the probability distribution ${\cal P}(U)$ 
and then evaluate
\begin{equation}
\langle {\cal O} \rangle = \lim_{N\rightarrow \infty} 
\frac{1}{N}\sum_{i=1}^N {\cal O}(U_i,\frac{1}{D(U_i)}) \ .
\label{eq:average}
\end{equation}
At finite $N$, the estimate of ${\cal O}$ is approximate, with an uncertainty
that can be shown to scale as ${\cal O}(1/\sqrt{N})$.

Both for the gauge
field configuration generation and the evaluation of
Eq.~(\ref{eq:average}), the linear system of equations,
\begin{equation}
 {D_m(U)}\chi =  \phi\,,
 \label{eq:linearsyst}
\end{equation}
must be solved, where $m$ is the quark mass
and the vectors $\chi$ and $\phi$  will be discussed below.  
Since $D(U)$ is sparse, iterative solvers can be used.  
The condition number of $D(U)$  (the ratio of largest to smallest eigenvalue),
and therefore
the computational resources required for  the 
solution of Eq.~(\ref{eq:linearsyst}),
is inversely proportional to
the quark mass.  
Since the physical quark masses for the up and down quarks are  small
relative to the typical scale of QCD, 
$D(U)$ has a large condition
number, and it is only recently that calculations at the physical quark masses have become computationally feasible.
The vast majority of the computational resources used in
nuclear physics LQCD calculations 
are devoted to the solution of this linear system,
both in the context of gauge-field generation and in the later stage
of the calculation of physical observables through
Eq.~(\ref{eq:average}). Significant gains in the efficiency of these calculations
have been achieved through applications of 
state-of-the-art
numerical linear algebra algorithms such as 
{\it deflation}~\cite{Morgan:2004zh,Darnell:2007dr,Stathopoulos:2007zi,Luscher:2007es}, 
and 
{\it multigrid}~\cite{Brannick:2007ue,Babich:2010qb,Frommer:2013fsa},
as will be further discussed  below.

\subsection{Gauge Field Generation: Hybrid Monte Carlo}

\noindent
The first stage of LQCD calculations is the
generation of suitable ensembles of gauge configurations. 
At present, the most efficient
algorithm for generating such ensembles is Hybrid Monte Carlo
(HMC)~\cite{Duane:1987de}.  
Because of the quark determinant, methods
that rely on local updates of the fields, such as the heatbath or the
Metropolis algorithms, are of limited use 
as their computational requirements  scale poorly
with the volume,
${\cal O}(V^2)$. 
HMC involves a
noisy representation of the determinant and introduces global updates
of the gauge fields, achieving volume scaling of ${\cal O}(V^\alpha)$ where
$\alpha\sim 1$. 
Other methods, such as the $\Phi$-algorithm and
the $R$-algorithm~\cite{Gottlieb:1987mq}, have also been used, however,
such methods are not exact and have a small systematic error that
must be carefully controlled.  
The global update of the gauge field using HMC is
obtained through a Hamiltonian evolution from an initial gauge-field
configuration and random initial
momenta (drawn from a Gaussian distribution).
In order to integrate the Hamiltonian dynamics, 
reversible discrete integrators are used 
so that detailed balance of the update procedure is maintained, as is required for Eq.~(\ref{eq:average})  to be satisfied. 
The simplest forms of such integrators are of  second order, however recently, higher order integrators have been 
employed following the work of Omelyan {\it et al}.~\cite{Omelyan:2003}.
The volume scaling of the algorithm~ \cite{Bernard:2002pd,Ukawa:2002pc,Urbach:2005ji,Kennedy:2006ax}  depends on the integrator,
with the resource requirements, $C$, scaling as
\begin{equation}
C = K \left( \frac{m_\pi}{m_\rho}\right)^{-z} V^{1+1/2n} \frac{1}{a^7}\,,
\label{eq:cost}
\end{equation}
where $n$ is the order of the integrator, 
$m_\pi$ is the pion mass, 
$m_\rho$ is the $\rho$-meson mass
(with $ \frac{m_\pi}{m_\rho} \propto \sqrt{m_q\over\Lambda_{\rm QCD}}$), 
$V$ is the volume
of the system, $K$ is a constant of appropriate dimensions, and $z$ is
an exponent that ranges between 4 and 6. 
Currently, 
several algorithmic improvements, 
such as preconditioned HMC~\cite{Hasenbusch:2001ne,Hasenbusch:2002ai,Luscher:2005rx}, are often used to reduce 
 $K$ and $z$,
and further,  higher order integrators  result in  better volume scaling.

\subsection{Continuum Limit and Autocorrelations}
\noindent
There are a large class of lattice gauge actions 
that have QCD as their continuum limit. 
For that reason, a variety of lattice actions are  used by different collaborations
worldwide. 
Comparing the continuum limit results obtained for a given quantity 
using different actions provides confidence that the calculations are 
performed correctly and uncertainties are quantified appropriately.

Continuum extrapolated results in the isospin limit are  functions
of only three parameters 
(four including strong isospin breaking), 
the values of the quark masses (up/down and strange)
and the characteristic QCD scale, $\Lambda_{\rm QCD}$
that emerges from quantum effects.
The tuning of the quark masses can be performed so that
chosen meson masses coincide with their experimental values. 
In the recent
years, several calculations have used the ratios
\begin{eqnarray}
l_\Omega & = &  \frac{m_\pi^2}{2 m_\Omega^2}\;\;\; 
{\rm and}\;\;\; 
s_\Omega \ =\  \frac{2 m_K^2 -  m_\pi^2}{m_\Omega^2}
\ \ ,
\label{eq:tuningclover}
\end{eqnarray}
to tune the bare quark masses,
where $m_\pi$ and $m_K$ are the  (isospin-averaged) pion and
kaon masses, respectively, and $m_\Omega$ is the
mass of the  $\Omega$ baryon~\cite{Lin:2008pr}.
By demanding that these ratios reproduce
their experimental values, 
the bare light- and strange-quark masses in the
calculation(s) are determined. 
The continuum limit can be taken  keeping
these ratios fixed.  
In addition, $\Lambda_{\rm QCD}$, or equivalently the
inverse lattice spacing $a^{-1} $, 
is determined in physical units
(MeV) using the experimental value of another hadron mass,
or a derived quantity such as the Sommer parameter~\cite{Sommer:1993ce} or the 
$w_0$ parameter~\cite{Luscher:2010iy,Borsanyi:2012zs}. 
The mass of the $\Omega$ baryon is currently a popular choice
as it depends weakly on the up- and down-quark masses.
Provided that the matching to experiment can be
performed at the physical quark masses, the scale determination is  robust,
with different choices of quantities with which to match resulting in only small changes in the extracted scale 
which can be quantified.
For a recent review on scale setting issues, the reader is referred to~\cite{Sommer:2014mea}.

Uncertainties introduced by the choice of the hadronic observable for scale setting arise from the discretization,
which is removed once the continuum limit is taken. 
Because such uncertainties percolate through to all computed observables, 
careful thought is needed in choosing the 
hadronic observable that sets the scale in order to minimize uncertainties in other physical quantities. 
A good choice of an observable is one that can be computed with  the smallest systematic and statistical uncertainties.   
The robustness of the  continuum extrapolation is improved if the scale setting observable is chosen to have only weak dependence on the discretization
and the light-quark masses.
The remaining systematic uncertainties from  setting the scale arise from physics that is not included in the calculation,
for instance,  electromagnetic (EM) effects,
isospin breaking effects, as well as the omission of vacuum polarization effects due to heavy flavors. 
All these effects are currently analyzed and steps are taken to minimize their impact. 
Calculations that include both isospin breaking 
and EM are under way~\cite{Drury:2013sfa,deDivitiis:2013xla,  Borsanyi:2014jba,Bazavov:2014wgs},
as are 
calculations with dynamical charm quarks~\cite{Bazavov:2014wgs,Bazavov:2012xda,Sommer:2010ui,Chiarappa:2006ae}.

For the continuum extrapolation, calculations at fine lattice spacings, $a\ll 1/\Lambda_{\rm QCD}$,
are required. 
Unfortunately, as the continuum
limit is approached, the autocorrelation time
(corresponding to the number of updates in the Markov chain of gauge configurations after which an observable is statistically independent)
becomes large for some observables . 
This was first observed in quenched calculations~\cite{Aoki:2002vt,Aoki:2005ga}, and  recently studied in detail in
dynamical fermion calculations~\cite{Schaefer:2010hu,Bruno:2014ova}. 
It was found that  for the topological charge, 
the integrated autocorrelation
time $\tau(Q^2)\sim a^{-5}$. 
This observation further
increases the 
computational resource requirements of
calculations near the continuum limit.
However, the  use of open boundary conditions
may help to resolve the problem~\cite{Luscher:2012av,Luscher:2014kea}
(although other recent work~\cite{McGlynn:2014bxa}
suggests this may be optimistic).
Unreliable estimates of both the statistical and systematic uncertainties may arise from long  autocorrelation times. 
The effects can be treated with statistical analysis methods on ensembles derived from Markov processes of length much 
longer than the autocorrelation time. 
Such methods are well understood and are part of the standard  methodologies used in the LQCD community. 
For a careful discussion of autocorrelation effects the reader should consult Refs.~\cite{Luscher:2010ae,Wolff:2003sm}.

\subsection{Quark Propagators}
\label{sec:prop}

\noindent 
A second major ingredient in almost all LQCD calculations 
is the {\it quark propagator}, $S(U)$, which is given by the inverse of the Dirac operator. As seen in 
Eq.~(\ref{eq:CorFunc}), after integrating out the 
quark degrees of freedom in the functional integrals
defining the correlation functions that need to be studied to extract physical observables,  an 
expression involving an integral over the remaining gauge degrees of freedom that depends on 
$S(U)\equiv D(U)^{-1}$ remains. 
The propagator
is a $12V\times 12V$ matrix in spacetime and color  and Dirac space, 
and 
each column
is the solution of the equation
\begin{equation}
[D(U)]_{X,Y} [S(U)]_{Y, X_0}  = G_{X,X_0}
\label{propEq}
\end{equation}
where G is a {\it source} that may be a Dirac delta function at a particular site, or a smeared  version that has
support in a particular region. 
In addition, momentum plane waves in fixed gauge, distillation and dilution vectors~\cite{Peardon:2009gh}
and various other structures can be used as sources.

Given the dimensions of the objects involved in Eq.~(\ref{propEq}) (for current large-scale calculations, $D(U)$ 
may be $\sim 10^{10}\times 10^{10}$), and the sparsity pattern of the Dirac matrix, iterative methods provide
the only practical approach to solving this linear system and determining the necessary components of the 
propagator. Most modern calculations use Krylov-space based solvers such as {\it conjugate gradient} (CG), 
{\it stabilised bi-conjugate gradient} (BiCGSTAB), 
or deflated versions such as {\it EigCG}~\cite{Stathopoulos:2007zi} for this task. Some discretized quark actions (e.g., the Wilson action) 
are such that the Dirac operator is not Hermitian, in which 
case many of the simplest algorithms must be applied to the {\it normal equations}, $[D^\dagger D] S =  D^\dagger G$,
instead of the direct system, with a resulting increase in computational resource requirements.
In recent years, {\it preconditioners} that reduce the condition number of the matrix
 have become quite common. 
One such preconditioner, known as {\it algebraic multigrid}  has been shown to
 be particularly efficient for the QCD problem~\cite{Frommer:2013fsa,Babich:2009pc}.
 Solution of these  linear systems forms a large part of the computational cost of LQCD calculations, and thus
 the calculations of these matrix inverses have been highly optimized for many computing architectures. In particular, 
optimised codes exist on IBM BlueGene and Cray supercomputers, clusters, {\tt nVidia} GPUs~\cite{Clark:2009wm,Babich:2011np,Winter:2014dka} 
and {\tt Intel} Xeon Phis.

For gauge-field configurations that are physically large compared to the QCD scale, $L \gg \Lambda_{\rm QCD}^{-1}$, it is beneficial to
make use of translational invariance  to exploit 
the full statistical power of the computationally expensive gauge-field configurations. 
To this end, it is
advantageous to compute propagators from multiple, physically-separated source locations on the same configuration,
and subsequently average measurements over these different locations. 
This amounts to solving Eq.~(\ref{propEq}) repeatedly with the same $D(U)$, 
but multiple different right-hand sides. 
This makes the application of more
complex solvers computationally viable; for solvers such as {\it EigCG} and {\it multigrid}, there is a significant setup cost involved that must be performed 
once but can then be reused to accelerate subsequent solves. 
By amortizing over a large number of solves, 
these algorithms lead to order-of-magnitude increases in computational speed compared to 
{\it CG} and even 
{\it BiCGSTAB}.

Since the methods used are iterative, applying a set of steps repeatedly until a convergence criterion is 
satisfied, the desired criterion and precision goal must be specified. 
Since there are significant  fluctuations	
intrinsic to the importance sampling of the gauge field, 
it is only useful to solve the above linear systems  to a 
precision that is marginally better than the gauge-field noise. 
However, for typical calculations at the present time, the desired precision
of solves is typically a relative error on the norm $||DS-G||<10^{-10}\sim10^{-12}$, 
approaching machine precision. 
While the final solution may be required with double precision accuracy, it is possible to obtain this accuracy in
a computationally expedient way by first solving the system in single precision and then using this solution as a starting 
point for the more expensive double precision solve, 
which will then converge in
relatively few iterations. On GPUs, half-precision
numerical representations are also available and result in effectively twice the performance of single precision.
The {\tt QUDA} 
library  for propagator inversions on GPUs
takes advantage of this feature and includes mixed half-double precision solves.

An interesting development in recent years has been the construction of improved estimators for many physical observables. 
These techniques aim to perform a modified set of measurements in which statistical fluctuations are small. 
A number of variants
to  this approach exist, 
such as low-mode averaging~\cite{Giusti:2002sm,DeGrand:2004qw} 
or truncated solver methods~\cite{Bali:2009hu},
which are being tested in single hadron calculations. One promising technique is covariant approximation 
averaging~\cite{Blum:2012uh,Shintani:2014vja}.
These methods attempt to speed up the solution of the linear systems by calculating low eigenmodes exactly 
or by performing ``sloppy precision solves'' and then correcting for the residual. 
They are currently being used for calculations of nucleon form factors and related observables.

Many observables, such as flavour singlet meson masses and iso-singlet matrix elements, 
 involve quark-line-disconnected contributions in which a quark propagator must be evaluated from
every site in the lattice to itself. 
Each solution of Eq.~(\ref{propEq}) provides a propagator from a single site to everywhere, 
so in order to have propagators from every site to itself, ${\cal O}(V)$ solutions of 
Eq.~(\ref{propEq})
are required. 
This is a prohibitive task to perform deterministically, but significant progress has been made in  
estimating such all-to-all propagators using stochastic volume sources.
Recent progress has also been
made with the introduction of probing methods~\cite{Stathopoulos:2013aci,Endress:2014qpa}.
Nevertheless, quantities that require all-to-all
propagators  remain challenging.

\subsection{Mistuning Input Parameters}

The tuning of the quark masses is accomplished by performing a combination of extrapolations from heavier masses 
and of low-statistics calculations in the vicinity of the parameter set of interest, followed by an interpolation to the desired point.
As this is accomplished with a relatively small number of measurements compared to those involved in the actual production, the tuning will always be imperfect.
Only after the production is complete,
involving calculations at multiple  lattice spacings and multiple volumes,
are the meson masses and scale setting known with high precision.
These can be used with great effect in subsequent tunings, 
however, the  mistunings require that 
small 
corrections are 
made to the results that have been generated in order 
to make predictions.
Consequently, multiple calculations are needed in the vicinity of the quark masses of interest in order to be able to systematically eliminate the impact of the mistuning.
Alternatively, reweighting methods~\cite{Liu:2012gm}
can be used to replace the determinant terms 
in Eq.~(\ref{eq:CorFunc}) 
with ones with corrected mass parameters.
As the quark masses will be close to the desired ones, and all physical results will be smooth functions (for sufficiently small deviations),
simple polynomial forms describing the behavior of the quantity of interest will be sufficient to interpolate to the desired quark masses.
For precision calculations, the uncertainty associated with this mistuning of input parameters must be quantified.

\section{Spectroscopy : Two-Point Correlation Functions}
\noindent
A central task of LQCD, is to perform hadron spectroscopy. 
To achieve this,  quark propagators are contracted together with the appropriate flavor, Dirac and spacetime structure 
to generate correlation functions with the desired quantum numbers.
These correlation functions  are then analyzed with an array of statistical techniques to extract energies and energy differences
and their corresponding uncertainties.

\subsection{Euclidean Space Correlation Functions}
\noindent
For lattice actions with a positive-definite transfer matrix~ \cite{Luscher:1976ms,Luscher:1984is},
such as the Wilson gauge and quark actions,
Euclidean space two-point correlation functions 
are the sums of exponential functions.~\footnote{
For many improved actions, terms in the action extend over multiple time slices and the transfer matrix is not positive definite at the lattice scale.
However,  a positive definite effective transfer matrix emerges over physical length scales. 
In addition, domain-wall fermions do not have a single time slice 4D transfer matrix,
and the
correlation functions can exhibit additional sinusoidally modulated
exponential behavior with a period set by the lattice
spacing. 
Until the continuum limit is taken, this introduces a systematic error that is difficult to quantify. 
Consequently, calculations that  seek to probe short distance details of QCD, 
such as the excited state spectrum, tend to use actions with no temporal improvement. 
If improvement is performed in this asymmetric way, this introduces a further systematic, in that the action is anisotropic -- spatial and temporal 
directions are not interchangeable. 
This translates into an anisotropy between spatial and temporal lattice spacings that must be determined 
and its impact on final uncertainties must be quantified.
}
The arguments of the exponentials 
are the product of Euclidean time with 
 eigenvalues of the finite-volume Hamiltonian.
 For a lattice that has infinite extent in the time direction, the
correlation functions become a single exponential at large times, 
dictated by the
ground-state energy and the overlap of the source and sink with the
ground state.  
As an example, consider the pion two-point function,
$C_{\pi^+}(t)$, generated from interpolating operators of the form
$\pi^+({\bf x},t)=\pi^-({\bf x},t)^\dagger
=\overline{u}({\bf x},t)\gamma_5 d({\bf x},t)$,
\begin{eqnarray}
\!\!\!
C_{\pi^+}(t) & = & 
\sum_{\bf x}\ \langle 0|\ \pi^- ({\bf x},t)\ \pi^+ ({\bf 0},0)\ |0\rangle
\nonumber\\
& = & 
\sum_n
{e^{-E_n t}\over 2 E_n}\  \sum_{\bf x}
\langle 0|\ \pi^- ({\bf  x},0) |n\rangle 
\langle n|\pi^+ ({\bf 0},0) |0\rangle
\ \rightarrow\ 
|Z_0|^2
\ {e^{-E_0 t}\over 2 E_0}
\  ,
\label{eq:singlepioncorrelator}
\end{eqnarray}
where the sum over all lattice sites at each time slice, $t$, projects onto 
 ${\bf p}={\bf 0}$  momentum states.
The source $\pi^+({\bf x},t)$ 
is not an eigenstate of QCD and not only couples
to single pion states, but also to
all other states with the quantum numbers of the  pion.
More generally, the source and sink 
can be distributed (smeared) over a subset of
lattice sites  to increase the overlap onto the ground state.
Eq.~(\ref{eq:singlepioncorrelator})  shows that the lowest energy-eigenvalue
extracted from the correlation function  corresponds to the mass of the
$\pi^+$ (and, more generally, the mass of the lightest hadronic state
that couples to the source and sink) in the finite volume.

Once such a correlation function has been calculated on a set of gauge-field configurations, 
the simplest  objective is to extract the argument of the exponential function that
persists at large times. One way to do this is to simply fit the correlation function
over a finite number of time-slices to a single exponential function.  
A second method, that is  
useful in visually assessing the quality of the calculation, 
is to construct the effective-mass (EM) function
from Eq.~(\ref{eq:singlepioncorrelator}) 
as
\begin{eqnarray}
M_{\rm eff.}(t; t_J) & = &
{1\over t_J}\  
\log\left({ C_{\pi^+}(t)\over  C_{\pi^+}(t+t_J)}\right)\ \rightarrow\ m_{\pi}
\ \ \ ,
\label{eq:effectivemassfunction}
\end{eqnarray}
where  $t, t_J$ and $M_{\rm eff.}(t; t_J)$ are in lattice units.
At large times, $M_{\rm eff.}(t; t_J)$ becomes  a constant equal to the mass of the
lightest state contributing to the correlation function~\footnote{
This is obviously the most simplistic approach to this problem.  
One well-known method to extract the ground state and excited state energies is
the variational method~\cite{Luscher:1990ck, Michael:1985ne},
which is discussed below.}. 
The anti-periodic (periodic) boundary-conditions (BCs)
in the time direction, imposed on the
quark (gluon)  fields, in order to recover the correct partition function, 
result in the  correlation functions being 
sums of forward and backward propagating exponentials
in the time direction.

More sophisticated methods aim to extract multiple energy eigenvalues.
Fitting correlation functions to the sum of $p$ exponentials 
(or hyperbolic  functions)
to extract the ground state energy
requires fitting ranges that start at time separations from the source
that are large enough so that the $p+1^{\rm th}$ and higher 
excited states make negligible
contributions. The determination of the minimum time separation that
can be included in the fit is sometimes subjective. 
Hence a systematic
uncertainty from the choice of the minimum time separation in the fit is
included,
and  is estimated by observing the variation of the
extracted results as a function of the choice of fitting interval.

Multi-hadron correlation functions are somewhat more complex, particularly with a finite temporal direction due to many allowed combinations of hadrons 
propagating backwards in time.  These require significantly more complex analysis ~\cite{Detmold:2008yn,Detmold:2011kw}.

\subsection{Scaling with Source Density and Number of Configurations}
\label{sec:SrcScaling}
\begin{figure}[!ht]
  \centering
  \includegraphics[width=0.48\columnwidth]{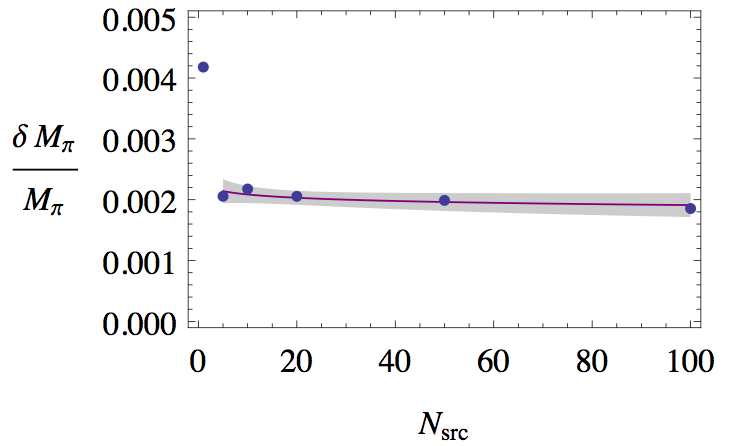}
  \includegraphics[width=0.48\columnwidth]{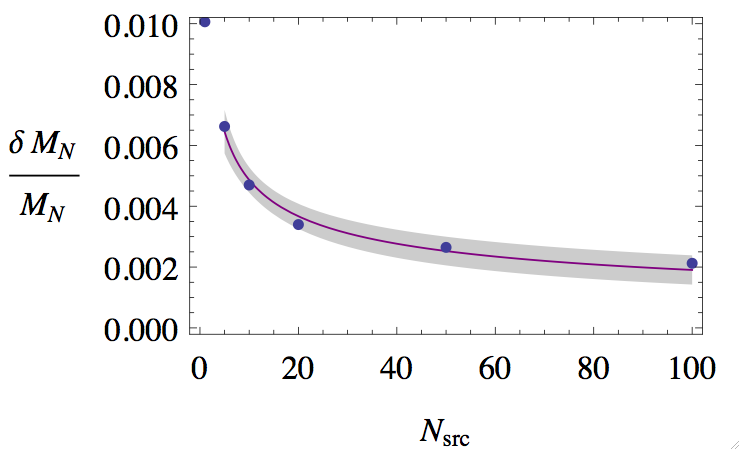}
  \caption{The fractional uncertainty in the extracted masses of the
    $\pi^+$ (left panel) and nucleon (right panel) as a function of the number of sources
    on each configuration~\protect\cite{Beane:2009kya}. Statistical and systematic
    uncertainties have been combined in quadrature. }
  \label{fig:sdvsNsrc}
\end{figure}

\noindent
It is interesting to explore the scaling of the uncertainties in the masses of the hadrons with 
the number of source locations and also with the number of gauge-field configurations.
On any given configuration, it is possible to perform a number of measurements equal to the number 
of lattice sites (for a given source structure).  
However, the uncertainty in the extracted energies is expected to 
scale as $1/\sqrt{N_{\rm src}}$
with the number of measurements, $N_{\rm src}$, at low source density, 
but when the density approaches one source per hadronic volume, 
deviations from this scaling are expected.
To demonstrate this behavior~\cite{Beane:2009kya},
the dependence of  hadron masses  on the number of sources 
obtained on an ensemble of  gauge configurations is shown  in
Figure~\ref{fig:sdvsNsrc}. 
The fractional uncertainties in
the masses of the $\pi^+$ and nucleon are shown as a function of the
number of sources used on each configuration. 
A simple fit of the form $\delta M/M = A N_{\rm src}^b$
returns exponents $b=-0.03(2)$ and  -0.41(3) for the
$\pi^+$, and nucleon, respectively.
The uncertainty in the energy of the pion is seen to saturate at relatively low source density compared to that of the heavier hadrons.
This behavior is expected from the differing Compton wavelengths of  the hadrons.
In contrast,
the scaling with the number of configurations is seen to be consistent with 
$1/\sqrt{N_{\rm cfg}}$
for each of the hadrons, as expected.
\begin{figure}[!ht]
  \centering
  \includegraphics[width=0.48\columnwidth]{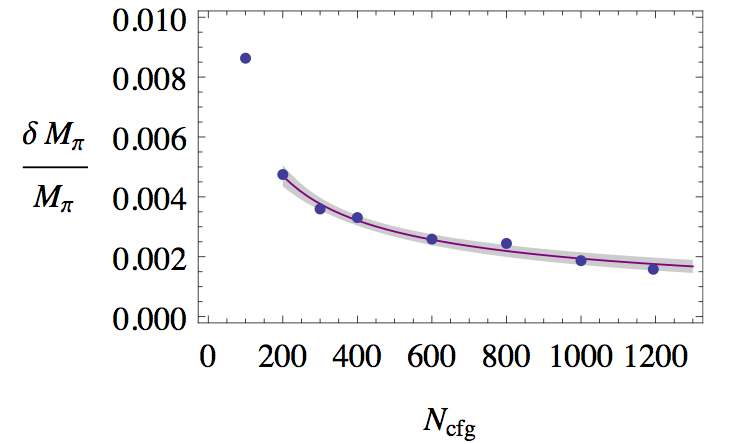}
  \includegraphics[width=0.48\columnwidth]{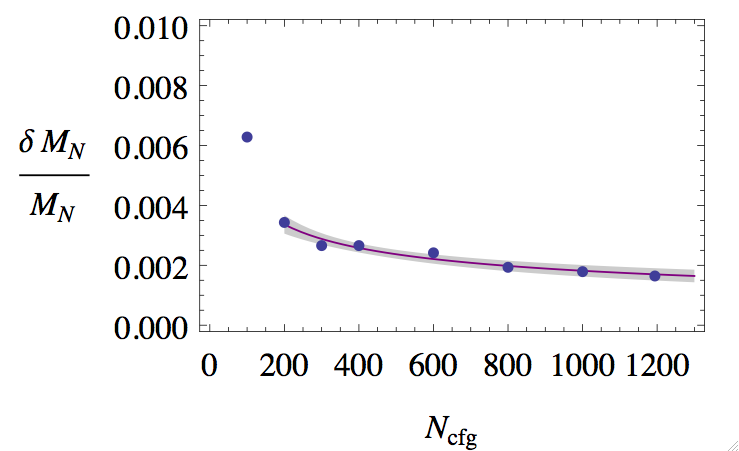}
  \caption{ The fractional uncertainty in the extracted masses of the
    $\pi^+$ and nucleon as a function of the number of gauge-field
    configurations~\protect\cite{Beane:2009kya}.
    Statistical and systematic
    uncertainties have been combined in quadrature. 
  }  \label{fig:sdvsNcfgB}
\end{figure}
Figure~\ref{fig:sdvsNcfgB} shows the fractional uncertainty in the mass
of the $\pi^+$ (left panel) and nucleon  (right panel) as a function of the number of
configurations.  An extrapolation can be performed with a fit to the
uncertainties in Figure~\ref{fig:sdvsNcfgB} of the form $\delta M/M = A
N_{\rm cfg}^b$.  The exponents extracted in these fits are -0.55(4)
and -0.38(4) for the $\pi^+$ and nucleon,
respectively~\cite{Beane:2009kya}.

\subsection{Analysis of Correlation Functions}

\subsubsection{Statistical Analysis Methods}
\label{sec:StatAn}
Since Monte-Carlo integration is used to compute the correlation
functions, they are subject to statistical uncertainty that must be
carefully determined.  
The main observables extracted from the
calculations presented in this review are energy eigenvalues and their
differences, which contain information about phase shifts, scattering
lengths and three-body interactions.  
The energy
eigenvalues are extracted by fitting the relevant correlation functions to a
sum of exponentials.  The optimal
values for the energy are extracted from correlated
$\chi^2$-minimization fits that take into account 
the correlations in the lattice calculations, 
both between different gauge configurations and between different times in a given configuration.
In the case of energy
shifts, the correlations between the energies of different states are also accounted for.  In
particular, the relevant parameters, such as the energies and the
amplitude of each state that contributes to the correlation function,
are determined as those that minimize
 \begin{equation}
\chi^2(A) = \sum_{i,j >i_{\rm min}}^{i_{\rm max}} 
\left[\bar G(t_i) - F(t_i, A)\right] \left[C^{-1}\right]_{ij} 
\left[\bar G(t_j) - F(t_j, A)\right]
\ \ \ ,
\end{equation}
where 
$F(t,A)$ is the fitting function, $A$ denotes the set of
fitting parameters over which $\chi^2(A)$ is minimized, and 
\begin{equation}
\!\!\!\!\!\!\!\!\!\!\!\!\!\!\!\!\!\!{}\bar G(t) = \frac{1}{N} \sum_{k=1}^N G_k(t)
\ \ ,\  \ 
C_{ij} ={1 \over
N(N-1)} 
 \sum\limits_{k=1}^N \left[G_k(t_i) - \bar G(t_i)\right] 
\left[G_k(t_j) - \bar G(t_j)\right] 
\ , \qquad\qquad\qquad\qquad\qquad\qquad
\end{equation}
are the average correlation function and correlation matrix, respectively.
The uncertainties in the fitted parameters are determined by the
boundaries of the ellipsoid defined by a given confidence
level~\footnote{For a pedagogical presentation of fitting see 
Ref.~\cite{DeGrand:1990ss}.},
typically $68\%$ or $90\%$.
It is important to
account for correlations in a manner that gives the best estimate of
the statistical uncertainty.  
Given $N$ independent measurements of
an energy level, it is straightforward to obtain the sample mean and
variance and thereby give an unbiased measure of the uncertainty in
the mean. 
However, in computing scattering parameters, the procedure
for determining the statistical uncertainties is somewhat more
involved because   the  relation between the
scattering amplitude and the energy levels of the two hadron
system is highly nonlinear. 
First, one is interested in the energy differences between the
energy levels of the two hadron system and the sum of the masses of
the two free hadrons (similarly for the case of three or more
hadrons).  
These energy differences can be determined in various ways.
The simplest is to perform fits to correlation functions of the
multi-hadron system and the single hadron system(s) generated on the
same gauge-field configurations and to form correlated differences of the
extracted energies.  
The ratios of correlation functions can also be analyzed,
where Jackknife and Bootstrap resampling methods are used to
determine the covariance matrix and then a correlated $\chi^2$-fit is performed~\cite{DeGrand:1990ss,Efron:1982}.

Beginning with a sample of $N$ elements, single-elimination Jackknife
removes the $k^{\rm th}$ element, leaving a sample of $N-1$ elements.  
Taking $R_k$ to be the desired ratio of correlation functions computed with
the $k^{\rm th}$ sample omitted from the full ensemble and $\bar R$ its
ensemble average,
the covariance matrix of the ratio of correlation
functions is  given by
\begin{equation}
C_{ij} = \frac{N-1}{N}  \sum_{k=1}^N \left[R_k(t_i) - \bar R(t_i)\right] 
\left[R_k(t_j) - \bar R(t_j)\right] \ .
\end{equation}
The Bootstrap method 
is a generalization of Jackknife.
Again, beginning with a sample of $N$ elements, in its simplest
implementation Bootstrap forms a new sample of $N$ elements
by randomly choosing values from the original sample, with
repetitions allowed. This procedure is repeated $N_B$ times
and a statistical analysis is carried out on each of the
Bootstrap ensembles. Now denoting $R_k$ as the $k^{\rm th}$  Bootstrap ensemble,
the covariance matrix is
estimated by
\begin{equation}
C_{ij} = \frac{1}{N_B-1}  \sum_{k=1}^{N_B} \left[R_k(t_i) - \bar R(t_i)\right] 
\left[R_k(t_j) - \bar R(t_j)\right] \ ,
\end{equation}
where now
\begin{equation}
\bar R \ =\ \frac{1}{N_B}  \sum_{k=1}^{N_B} R_k \ .
\end{equation}
The value of $N_B$ should be large enough so that stable and accurate
estimates are obtained. 
In computing the mean and uncertainty, both Jackknife and
Bootstrap lead to comparable results, as the distributions of correlation functions  are smooth.

\subsubsection{ Fitting Methodology}
\label{sec:Fitting}

Fitting correlation functions to a sum of
exponentials is a difficult problem. 
It is significantly simplified if
only the lowest energy eigenvalue is required. 
However, a great deal
of spectral information about QCD resides in the energy levels
above the ground state. 
In order to reduce the systematic uncertainty from excited-state contamination
at small time separations, it is important to have a
signal at large Euclidean times.  
However, at large times, 
the
statistical uncertainties in most correlation functions grow
exponentially and therefore the extraction of the lowest energy
eigenvalue at large times typically results in large
statistical uncertainties.  
In principle one can trade
statistical uncertainty growth for systematic uncertainty reduction by
developing improved sampling methods to reduce statistical
uncertainties in the correlation functions.
However,  in practical LQCD
calculations, one extracts as much information as possible from the
correlation functions at short times where the statistical
noise does not overwhelm the signal, but
multiple exponentials contribute to the correlation
functions.

Although the general multi-exponential fit problem is
difficult and not well behaved, systems of correlation functions can
be designed in order to optimize these fits.  Variational analyses on
symmetric positive-definite matrices of correlation functions have
been successfully used in the LQCD community to extract the
energy eigenvalues contributing to the correlation functions. 
These methods were originally introduced in
Refs.~\cite{Luscher:1990ck, Michael:1985ne}, and have been
subsequently
developed~\cite{Dudek:2009qf,Foley:2007ui,Basak:2007kj,Blossier:2009kd}.

\begin{figure}[htb!]
 \centering
 \includegraphics[width=0.85\textwidth]{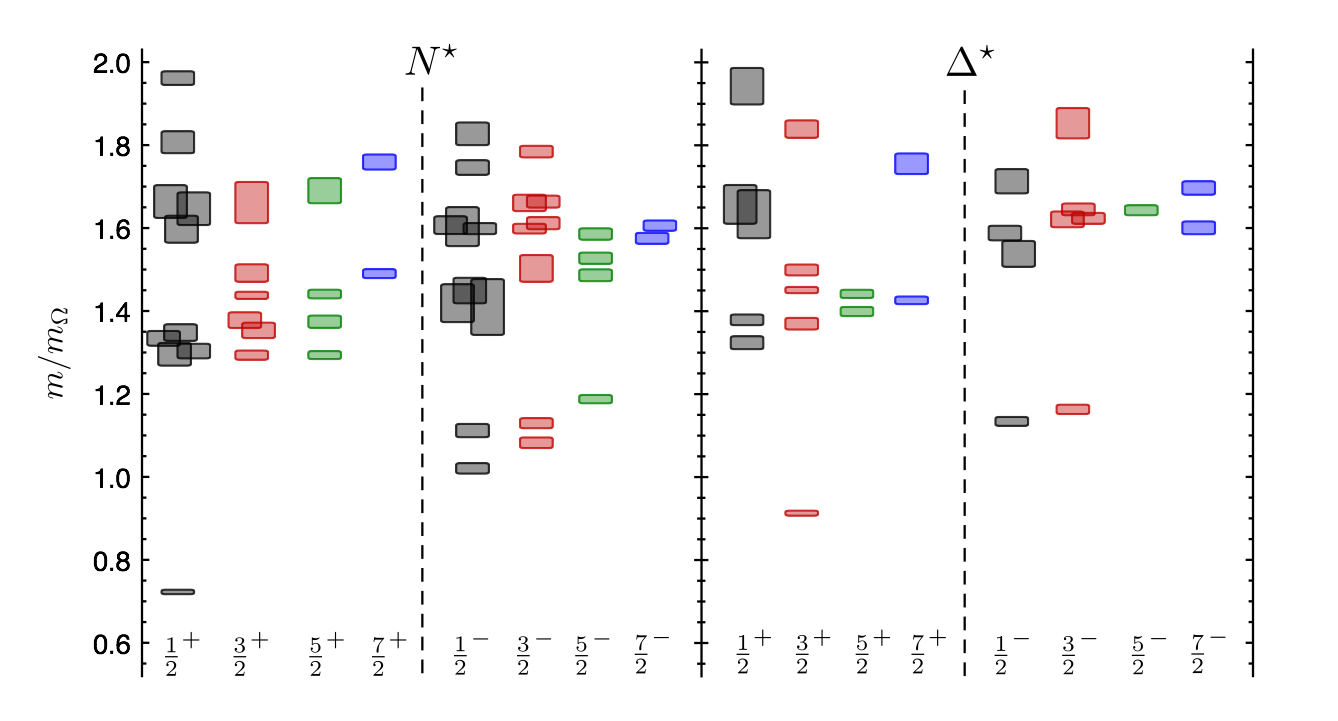}
 \caption{
 Excited nucleon spectra calculated at a pion mass of $m_\pi\sim 390~{\rm MeV}$~\protect\cite{Edwards:2012fx}.
 The solid regions correspond to the $68\%$ confidence intervals for the energies of the states.
          [Figure reproduced with the permission of the Hadron Spectrum Collaboration.]
}
 \label{fig:excBaryon}
\end{figure}

Generalizing the pion correlation function of Eq.~(\ref{eq:singlepioncorrelator})
to 
a set of operators, $\{ {\cal O}_i \}$,
of commensurate quantum numbers, the correlation functions can be defined
\begin{eqnarray}
{\cal C}_{ij}(t) 
& = & \langle {\cal O}^\dagger_i(t) {\cal O}_j(0) \rangle\ =\ 
\sum_n
{e^{-E_n t}\over 2 E_n}
\langle 0|\ {\cal O}^\dagger_i(t)  |n\rangle 
\langle n| {\cal O}_j(0) |0\rangle \ .
\label{eq:gencorrelator}
\end{eqnarray}
At large times,  the correlation functions are dominated by the
first few exponentials, and $E_0$, $E_1$, ...  can be extracted by considering multiple correlation functions. 
If sufficient
resources exist to construct a basis of interpolating operators, 
the orthonormality of state vectors can be used to extract multiple
energies in a controlled manner~\cite{Luscher:1990ck, Michael:1985ne}.
This is achieved by solving the generalized eigenvalue problem of the form
\begin{eqnarray}
{\cal C}(t)\; v_n\ =\ \lambda_n(t){\cal C}(t_0)\; v_n \ ,
\label{eq:LW}
\end{eqnarray}
where the $\lambda_n$ and the $v_n$ are the principal correlators 
and eigenvectors, respectively. The utility of this method
stems from the observation that at large times, the principal correlators
satisfy
\begin{eqnarray}
\lambda_n (t)\ =\ e^{-E_n(t-t_0)}\left( 1\ +\ {\cal O}(e^{-|\Delta E|(t-t_0)})\right) \ ,
\label{eq:LWuncertain}
\end{eqnarray}
where $\Delta E$ is the gap between the level of interest 
and the $N+1^{\rm th}$ level, where $N$ is the rank of ${\cal C}_{ij}$.
There are many implementations of this so-called variational method. 
In Figure~\ref{fig:excBaryon},
we show  recent results for the nucleon excited spectrum from the Hadron Spectrum Collaboration~\cite{Edwards:2012fx}
determined using these methods.

In many cases, computational cost  precludes the construction of a  
basis of interpolating operators required for the variational method to be employed. 
In these cases, Matrix-Prony and other related methods~\cite{Beane:2010em} facilitate 
an extraction of the low-lying levels from a set of at least two correlation functions
with distinct operator source and sink structure.

\subsubsection{Non-Gaussian Fluctuations and  Lepage's Argument.}
\label{sec:NGLepage}
\begin{figure}[htb!]
 \centering
 \includegraphics[width=0.7\textwidth]{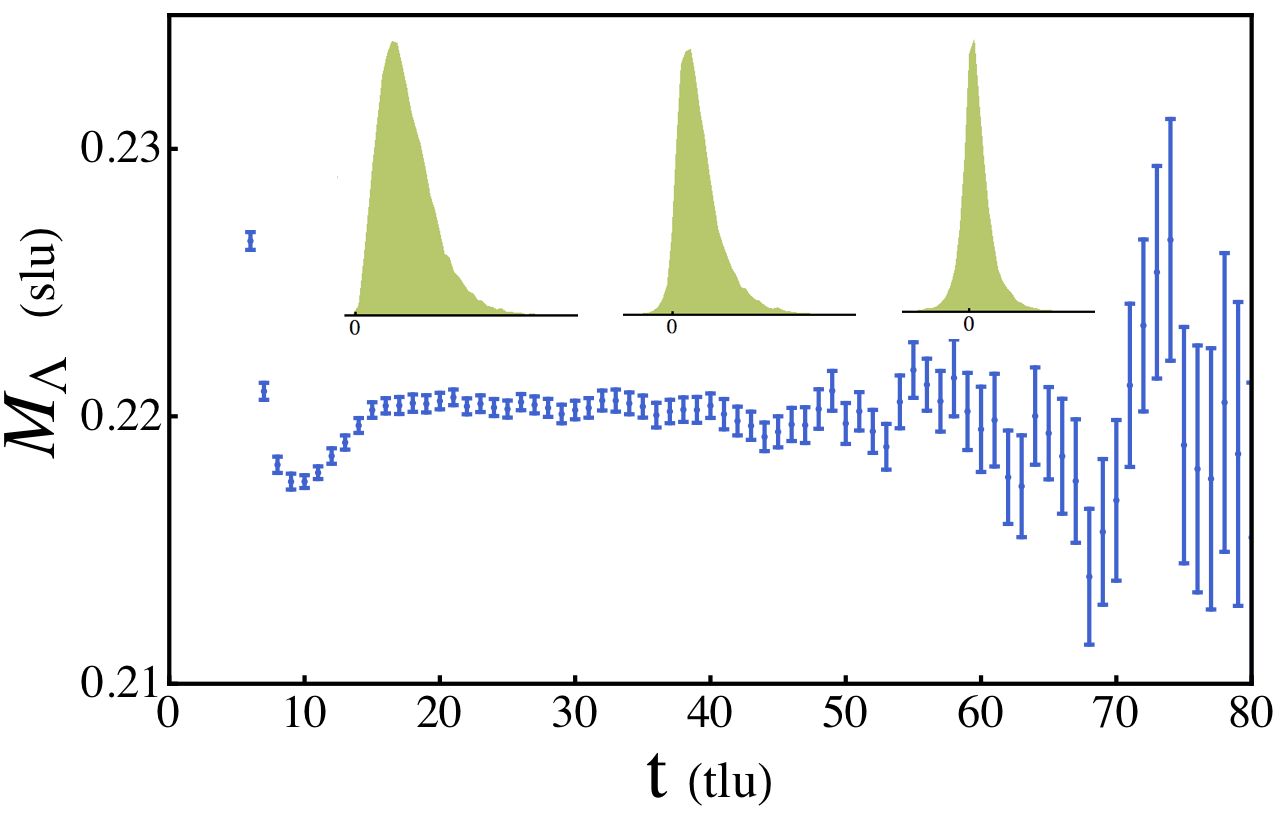}
 \caption{An effective mass plot (blue points) obtained from $\Lambda$-baryon correlation functions on an ensemble of 
 anisotropic clover gauge-field configurations.
 The time axis is in temporal lattice units (tlu), while the energy axis is in spatial lattice units (slu).
 The insets (from left to right)  show the (normalized) distribution of the correlation function at time 
 $t= 20, 40$ and $60$ tlu from the source. 
This is derived from  $\sim 120$ 
measurements on each of $\sim 700$ gauge-field configurations.  
These measurements were blocked down to 93 independent representative correlation  functions, 
and  Jackknife  was used to generate the  covariance matrix.
 }
 \label{fig:empdist}
\end{figure}

As QCD is a highly non-trivial interacting quantum field theory, the
quantum fluctuations of the quark and gluon fields, and hence in the
derived correlation functions, are non-Gaussian.  Following
Parisi~\cite{Parisi:1980ys}, Lepage~\cite{Lepage:1989hd} explored the
relation between the variance of a correlation function and its mean,
and in the process identified the exponentially degrading
signal-to-noise in baryonic systems.  This argument can be
generalized to arbitrary moments of any given correlation function.
Denote the single nucleon correlation function (projected to zero
momentum) as
\begin{eqnarray}
\langle\theta_N(t)\rangle & = & 
\sum_{\bf x}\ \Gamma_+^{\beta\alpha}\ \langle0|\ N^\alpha({\bf x},t) \overline{N}^\beta ({\bf 0},0)\ |0\rangle
\ \rightarrow\ Z_N\ e^{-M_N t}
\ \ \ ,
\label{eq:StoNsingle}
\end{eqnarray}
where $N^\alpha$ represents an interpolating operator for the nucleon, 
$\Gamma_+$ denotes a positive-energy projector, and the
angle brackets denote the statistical average over
calculations on an ensemble of gauge-field configurations.  
At short
times, for operators that have a large overlap with the nucleon ground
state, the moments of this correlation function are  dictated by
the multi-nucleon-anti-nucleon states, 
$\langle \left(\theta_N^\dagger  \theta_N\right)^n \rangle\sim e^{-n M_N t}$  
(neglecting the interactions between nucleons), and the distribution of
correlation functions is asymmetric with a non-exponentially degrading
signal-to-noise ratio~\footnote{
The magnitude of the suppression of
  lighter hadronic states with the same quantum numbers as the
  multi-nucleon and anti-nucleon state, as well as the same number of
  quark and anti-quark propagators, such as multi-pion states, depends
  upon the structure of the sources and sinks.  
  When the sink is
  momentum projected over the lattice volume, overlap onto the
  multi-pion states are suppressed by the ratio of the
  volume of the nucleon compared to the lattice spatial volume,
  delaying the onset of the exponential degradation of the
  signal-to-noise ratio.  
  }.  
At large times, however, the lightest
states with the appropriate quantum numbers dominate, and the moments
scale as
\begin{eqnarray}
\langle  \left(\theta_N^\dagger  \theta_N\right)^{2n} \rangle
& \sim & e^{-3 n m_\pi t}
\ \ ,\ \ 
\langle  \left(\theta_N^\dagger  \theta_N\right)^{2n+1} \rangle
\ \sim\ 
e^{-M_N t} e^{-3 n m_\pi t}
\ \ \ .
\label{eq:StoNlarget}
\end{eqnarray}
The odd moments degrade exponentially compared with the even moments, 
leaving a non-Gaussian, but symmetric, 
distribution with a mean  $\sim e^{-M_N t}$ and variance  $\sim e^{-3 m_\pi t}$.
In Figure~\ref{fig:empdist}, we show the effective mass of the $\Lambda$-baryon obtained at 
a pion mass of $m_\pi\sim 390~{\rm MeV}$ on an ensemble of anisotropic gauge-field configurations~\cite{Beane:2011pc}. 
The inset histograms in this figure show the distribution of correlation functions, from which the time evolution from an asymmetric 
distribution with a non-zero mean value, to a non-Gaussian symmetric distribution suffering from an exponentially degrading signal-to-noise ratio is clearly evident. 
At large times, 
exponentially large computational resources are required 
to precisely extract the mean value of a baryon correlation function, 
which has an uncertainty  $\sim  e^{-3  m_\pi t /2}/\sqrt{N}$ for a large number of measurements, $N$  
(where Gaussian statistics have been assumed to provide an estimate of the scaling of the uncertainty in the mean).
A closer inspection of the time 
dependence of the moments of the correlation functions shows that there is an intermediate 
time range, dictated by the structure of the sources and sinks used to generate the correlation function~\cite{Beane:2009kya},
in which the signal-to-noise ratio is degrading much less severely than at asymptotically large times.
This  ``Golden Window'' 
is  seen in the effective mass shown in Figure~\ref{fig:empdist} between $t\sim 15$ and $50~{\rm tlu}$.
Efforts have been made~\cite{Detmold:2014hla} to optimize the signal-to-noise ratio by 
combining optimizing the overlap of the source and sink onto the correlation function(s) of interest 
with minimizing the overlap onto the corresponding variance correlation function(s).

It is worth attempting to understand the mechanisms responsible for the statistical behavior of the nuclear 
correlation functions.  
At each point in spacetime, 
the relevant parts of 
a light-quark propagator can be encoded in a $12\times 12$ matrix in Dirac $\otimes$ color space.
Loosely speaking, in color space, for each pair of Dirac indices,
the pion correlation functions arise from the 
sum of the squared norms
of the columns of this matrix while the nucleon correlation functions  arise from its determinant.  
The elements of each column scale as 
$\sim e^{-{1\over 2} m_\pi t}$, while the determinants scale as $e^{- M_N t}$.  
As the norms and orientations of each of the columns of these matrices are fluctuating because of interactions 
with the gauge field, 
with average values that are  becoming linearly dependent, the  signal-to-noise problem arises.

\subsubsection{Blocking, the Central Limit Theorem and Robust Estimators.}

Correlation functions can be determined  from multiple  
source locations on a given gauge-field configuration.
These measurements are correlated with each other as they result from the same sample of gluon fields, and,
in general,  cannot be treated as statistically independent.  
Because the correlation functions  become translationally invariant after averaging,
these measurements 
can be averaged together (blocked) 
to generate one representative correlation function for that gauge-field configuration.
More generally, because of the correlation between  nearby 
gauge-field configurations produced in a Markov chain, 
quantified by analysis of auto-correlation functions, 
the measurements performed over multiple gauge-field 
configurations are typically blocked together to produce one representative correlation function from a ``patch'' of the Markov chain.  
For a long Markov chain, or multiple independent Markov chains, 
there will be a large number of independent representative correlation functions that, by the central limit theorem,
will have an approximately  Gaussian-distributed mean.
In general, this set of blocked correlation functions are analyzed using correlated $\chi^2$-minimization methods 
(see Sec.~\ref{sec:StatAn})
assuming Gaussian statistics.

As computational resources are limited,  only a finite number of measurements of each correlation function can be performed.
The underlying distributions for nuclear correlation functions 
are non-Gaussian with extended tails, 
as seen in Figure~\ref{fig:empdist},
and therefore 
outliers are typically present in any sample, 
which  can lead to poor convergence of the mean
(for a discussion of the ``noise'' associated with these and other such calculations, see Ref.~\cite{Endres:2011mm}).
Dealing with outliers of distributions is required in many areas beyond physics, and there is extensive literature on 
{\it robust estimators} that are resilient to their presence, such as the median or the Hodges-Lehmann (HL) estimator.
However, for the quantities we are interested in, it  is the mean value (vacuum-expectation value)
that is required, and  not the median or mode.
With sufficient sampling and 
blocking,  the mean of the distribution of any given correlation functions is expected to  tend to a  Gaussian distribution by the central limit theorem,
for which the mean, median and mode coincide.
\begin{figure}[htb!]
 \centering
 \includegraphics[width=0.65\textwidth]{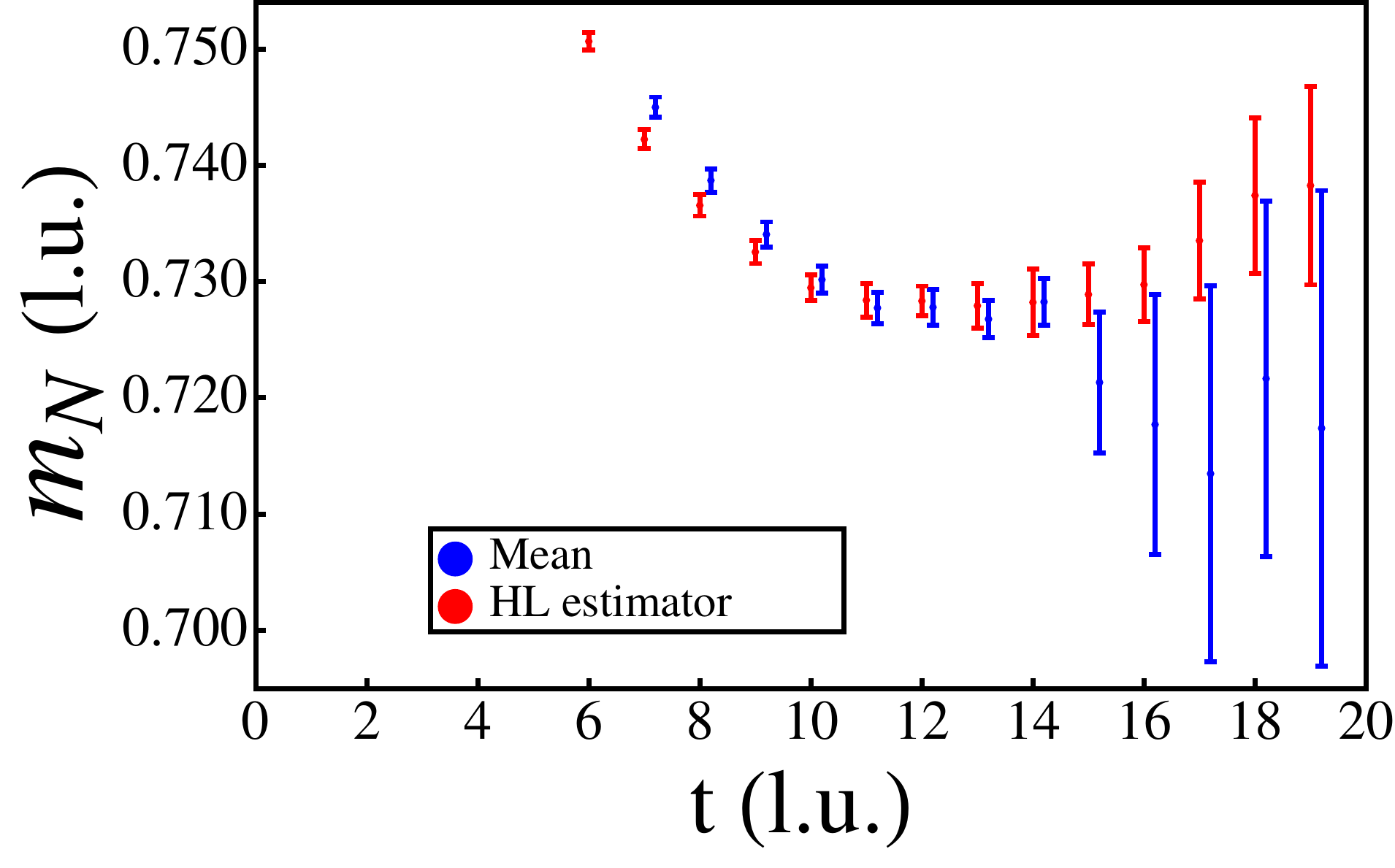}
 \caption{Effective mass plots for the nucleon obtained  from an ensemble of isotropic clover gauge-field configurations. 
 Blocked correlation functions are analyzed with the mean using Jackknife (blue) and Hodges-Lehman using Bootstrap (red).   }
 \label{fig:HLmean}
\end{figure}
As an example, 
the nucleon effective masses obtained from an ensemble of isotropic clover gauge-field configurations ($24^3\times 64$, $m_\pi\sim 430~{\rm MeV}$),
analyzed with both the mean under Jackknife  and Hodges-Lehman under Bootstrap, from $\sim 100$ blocked representative correlators,  
are shown in Figure~\ref{fig:HLmean}.
The non-Gaussian distribution present in the blocked correlation functions 
leads to a large estimated variance at late times,
while the  HL-estimator is more robust.
Investigations in this direction are ongoing.

\subsection{Observables}
\noindent
The energies and energy differences extracted from  correlation functions calculated on 
one ensemble of gauge-field configurations deviate from those of QCD, even in the infinite sampling limit, 
because of the finite lattice spacing and the finite lattice volume.
Further, there are also  deviations because of unphysical values of quark masses and/or   
imperfections in their tuning.

\subsubsection{Finite Lattice Spacing and the Continuum Limit.}

Because of the computational resource requirements, most calculations of quantities of importance for 
nuclear physics have been performed at one, in some cases two, and in very few instances three lattice spacings.
In contrast, simpler quantities, such as the meson decay constants, have been computed with 
precision at multiple lattice spacings and extrapolated to the continuum.
\begin{figure}[htb!]
 \centering
 \includegraphics[width=0.65\textwidth]{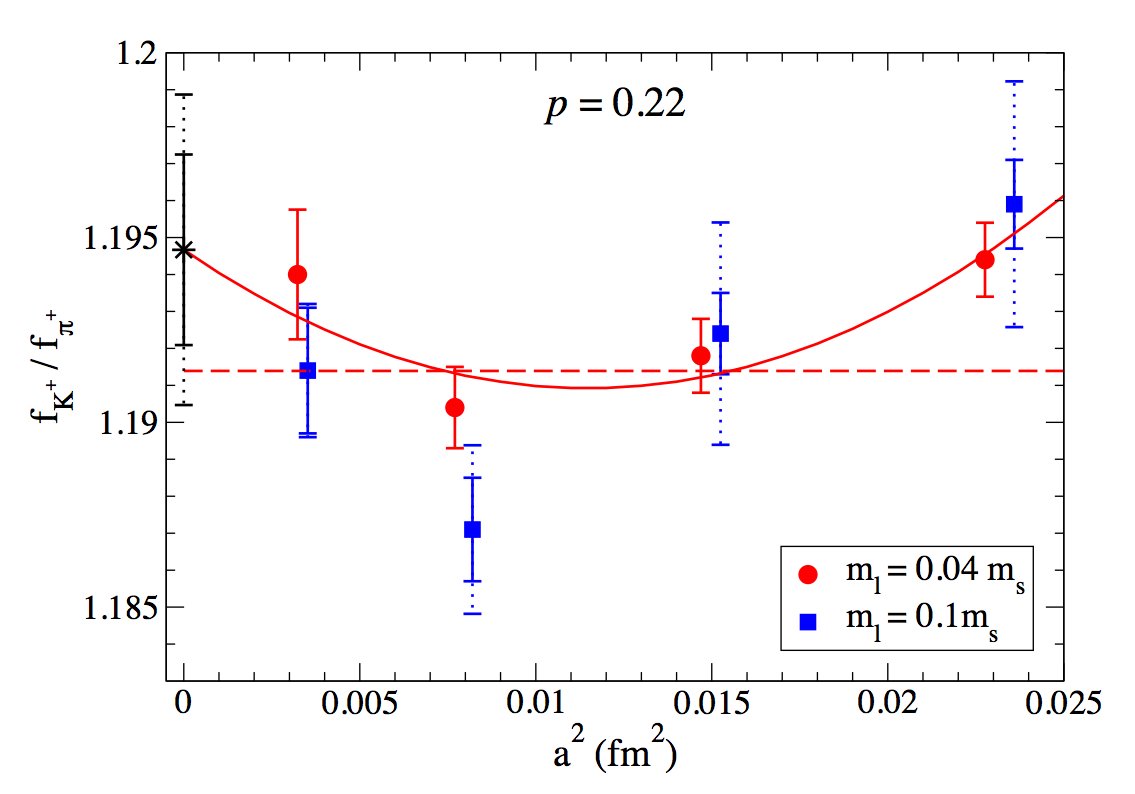}
 \caption{
The MILC collaboration's recent continuum extrapolation of $f_{K^+}/f_{\pi^+}$~\protect\cite{Bazavov:2013vwa}. 
 The solid (red) curve shows the result of fitting to the form $c_0 + c_2 a^2 + c_4 a^4$, while the dashed (red) line 
 corresponds to fitting a constant to the smallest two lattice spacings.
 The black point at $a=0$ corresponds to the continuum extrapolation, with the inner 
 uncertainty being statistical and the outer being the statistical and systematic uncertainties combined in quadrature.
 [Figure reproduced with the permission of the MILC collaboration.]
 }
 \label{fig:milcfkfpi}
\end{figure}
As an example,
the MILC collaboration's~\cite{Bazavov:2013vwa}
recent continuum extrapolation of 
the ratio of decays constants, $f_K/f_\pi$, determined at a pion mass of $m_\pi\sim 135~{\rm MeV}$ 
with $n_f = 2+1+1$ dynamical quarks, is shown in Figure~\ref{fig:milcfkfpi}.
MILC has produced large ensembles of gauge-field configuration using 
a one-loop Symanzik-improved gauge action for the gluons
and the HISQ (highly-improved staggered quark) action for the quarks 
with lattice spacings of $a\sim 0.06, 0.09, 0.12$ and  $0.15~{\rm fm}$, 
allowing for continuum extrapolations involving four independent points.~\footnote{
Due to the mistunings of quark masses, as discussed previously, such continuum extrapolations also require 
interpolations in the quark masses.
}
Two different extrapolations of their results, shown and described in Figure~\ref{fig:milcfkfpi}, 
provide consistent continuum limit values

The lattice-spacing dependence of observables can be determined from the Symanzik action~\cite{Symanzik:1983gh}, 
dictated by the symmetries of the discretized  action,
that describes the dynamics of the quarks and gluons at momentum scales much less than the 
inverse lattice spacing.  The operators in this EFT are formed from the quark and gluon fields with 
arbitrary numbers of derivatives and insertions of the quark mass matrices, with coefficients that scale with the 
appropriate power of the lattice spacing.
It is the matrix elements of these operators between the hadronic states of interest that dictate the lattice-spacing 
dependent deviations from  QCD.  
While the Symanzik
action lacks Lorentz invariance and rotational symmetry, it is constrained by the residual hypercubic symmetry
of the discretized action. 
The presently available computational resources have not permitted
calculations of sufficient precision to isolate lattice-spacing artifacts beyond 
polynomial in the lattice spacing,
however, 
such terms are present, for instance,
from the intrinsic logarithmic scale dependence of the coefficients in the Symanzik action.

\begin{figure}[!ht]
  \centering
  \includegraphics[width=0.49\textwidth]{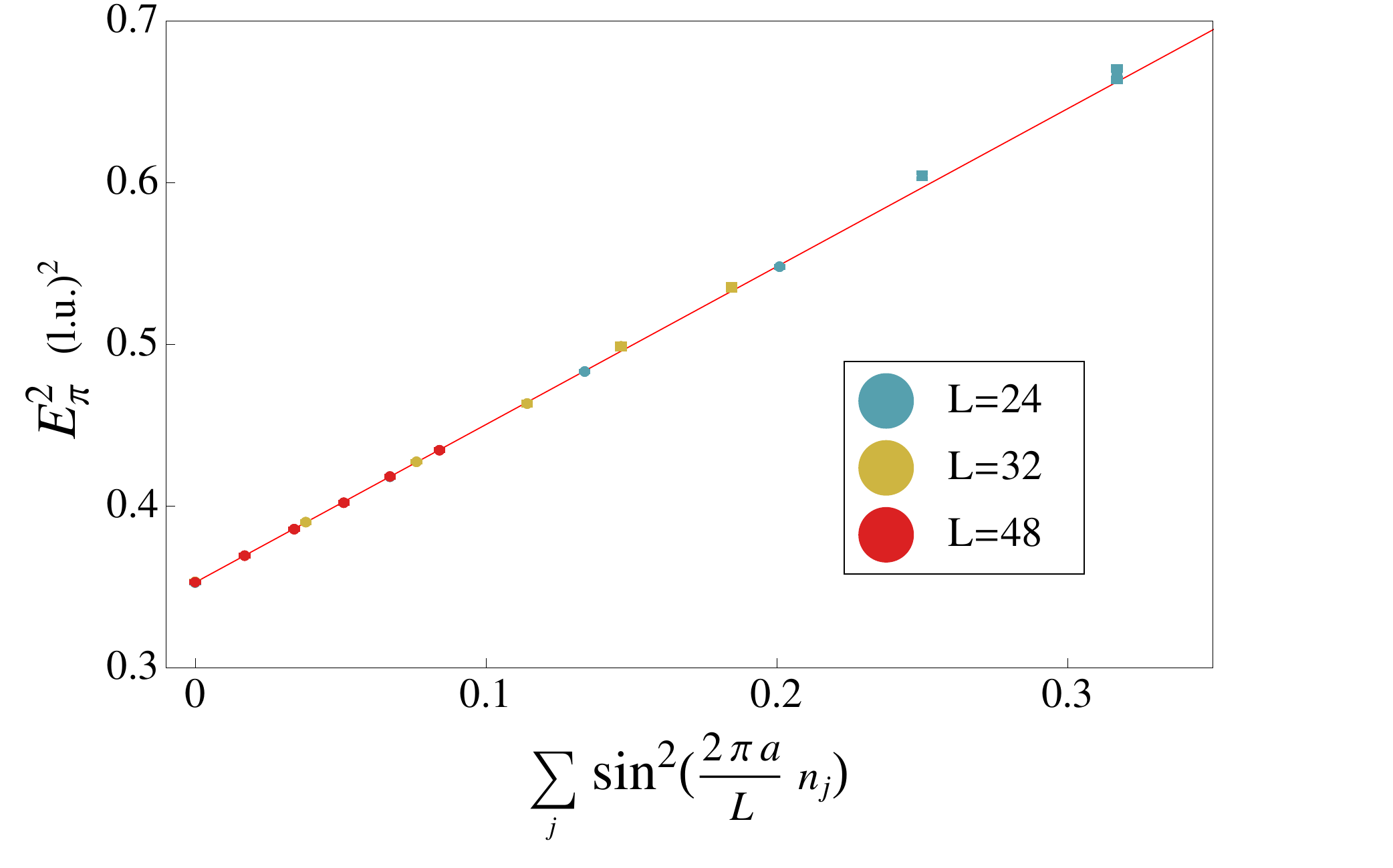}\
  \
  \includegraphics[width=0.49\textwidth]{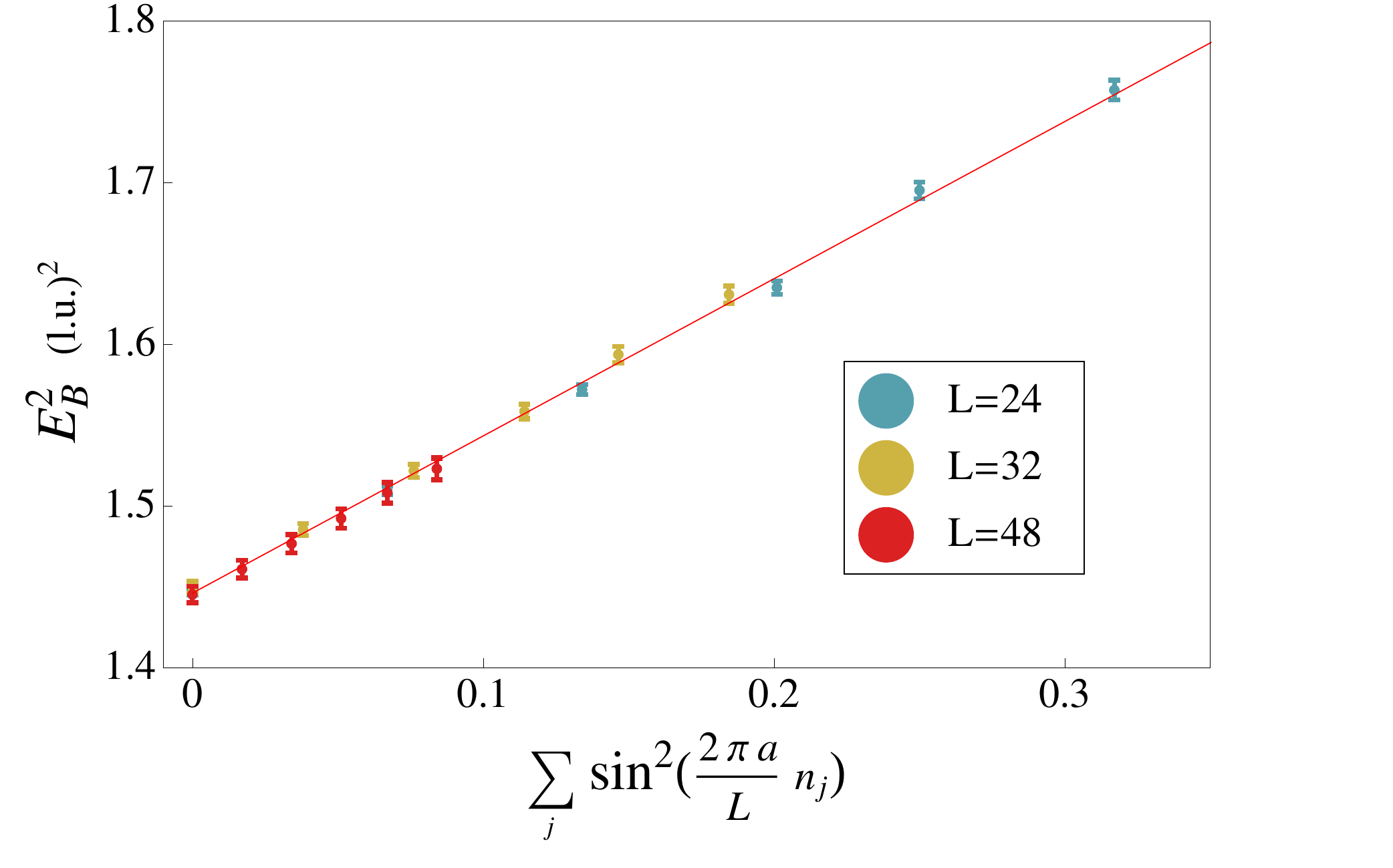}\
  \
  \caption{ Examples of the squared energy 
  (in $({\rm l.u.})^2$) of the single
    pion and nucleon as a function of $\sum\limits_j\
    \sin^2\left({2\pi a\over L} n_j\right)$
    at the flavor-SU(3) symmetric point~\protect\cite{Beane:2012vq,Beane:2013br}.  
    The figures show the results of LQCD calculations performed on isotropic clover gauge-field configurations,
    along  with the best linear fits.  }
  \label{fig:Dispersion}
\end{figure}

In order to have confidence in the extraction of multibaryon binding
energies and scattering phase shifts, 
and to be able to quantify one of the systematic
uncertainties in these determinations, it is important to determine
the single-hadron dispersion relations with precision.  
Example calculations of the energies of the pion and nucleon as a
function of $\sum\limits_j\ \sin^2\left({2\pi a\over L} n_j\right)$
are shown in Figure~\ref{fig:Dispersion}~\cite{Beane:2012vq,Beane:2013br}, 
where the triplet of integers ${\bf n}=(n_1, n_2,n_3)$ is related to the
 momentum of the state via 
${\bf P} = \left({2\pi\over L}\right) {\bf n}$.
In these LQCD calculations, the energy of the hadron can be related to its
lattice momentum through dispersion relations of the form
\begin{eqnarray}
  \left(\,a\,E_H\right)^2 
  & = &
  (a\,M_H)^2 
  \ +\
  \frac{1}{\xi_H^2}
  \ \sum_j\ \sin^2
  \left(\frac{2\,\pi\,a}{L} n_j \right) 
  \ +\  ...
  \ ,
  \label{eq:latticeDisp}
\end{eqnarray}
where  $\xi_H$ is the the anisotropy parameter  for hadron $H$ (or equivalently its fractional speed of light
$\beta_H=1/\xi_H$).
In the continuum limit, this reduces to $E_H^2 = |{\bf p}_H|^2 \beta_H^2  + M_H^2 $, as required.
At any finite lattice spacing, 
the relation between energy and momentum of any given hadron involves an infinite number of terms that 
respect the underlying hypercubic symmetry,
and this relation can only be determined by direct calculation.
The measured dispersion relation, and associated uncertainties, 
are necessary for determining multihadron binding energies and scattering phase shifts,
as will be discussed below.
Even after accounting for the lattice dispersion relation, 
these quantities
 will have residual dependence upon the lattice spacing 
because of modifications to the hadronic interactions, 
and an extrapolation in lattice spacing is required to obtain their continuum limit values.

A final comment regarding  finite lattice spacing concerns the recovery of rotational symmetry from the 
underlying hypercubic symmetry and the mixing of operators with those with lower 
angular momentum (and hence lower dimension).
It has recently been shown that in the limit of a lattice spacing that is small compared with any of the 
intrinsic length scales of the system, including the renormalization scale needed for certain operators,
the breaking of rotational symmetry scales as ${\cal O}(a^n)$ with $n\ge 2$, and its effects vanish in the continuum limit~\cite{Davoudi:2012ya}.
This results in  a parametric suppression of higher-dimension operators mixing with ones of lower dimension.

\subsubsection{Finite-Volume Effects and Boundary Conditions (I): Single Hadrons.}

For localized hadronic systems, such as single mesons, baryons and nuclei, if the objective of a LQCD calculation is 
to determine its mass or binding energy, then it is desirable to work in the largest practical lattice volume, both 
in the spatial and temporal directions. 
Ground-state energies, in situations where the lattice volume and temporal extent are much larger than the hadronic size,
have finite-volume  (FV) effects that scale exponentially with the lattice volume, a result that follows 
straightforwardly from considering the FV system in terms of its image systems~\cite{Luscher:1986pf,Hasenfratz:1989pk,Luscher:1990ux}.

For single mesons and baryons, calculations of the FV effects have been performed in the p-regime 
of chiral perturbation theory ($\chi$PT)
(the regime in which the momentum and quark mass divided by the chiral symmetry breaking scale
are the expansion parameters),
and are found to agree well with the results obtained in LQCD calculations at unphysical
light-quark masses, e.g. Refs.~\cite{Beane:2011pc,Colangelo:2006mp,Colangelo:2010ba},
but remain to be verified at the physical point.
However, for the pseudo-scalar mesons, the FV modifications to the decay constants and masses have been calculated beyond the 
one-loop level~\cite{Colangelo:2005gd}, and it has been shown that removing these FV effects from the results of 
LQCD calculations improves the overall fit quality at and near the physical point~\cite{Durr:2013goa}.
These works suggest that loop-level $\chi$PT calculations  describe the FV modifications to meson masses in large volumes as expected.
In using the low-energy EFT to determine the FV effects, it is implicit that the FV effects can be described 
by modifications of loop diagrams involving the lowest-lying mesons.
The reason this type of seperation is possible is because the FV modification to the coefficients of the local operators
in the low-energy EFT, in this case $\chi$PT, are exponentially suppressed by the size of the hadron, scaling as
$\sim e^{-L/r_\chi}$, where $1/r_\chi$ is set by the mass of the $\rho$ or the chiral symmetry breaking scale, $\Lambda_\chi$,
as opposed to 
parametrically larger
$\sim e^{-m_\pi L}$ 
behavior that results from the loop diagrams.

\begin{figure}[!ht]
  \centering
     \includegraphics[width=0.6\textwidth]{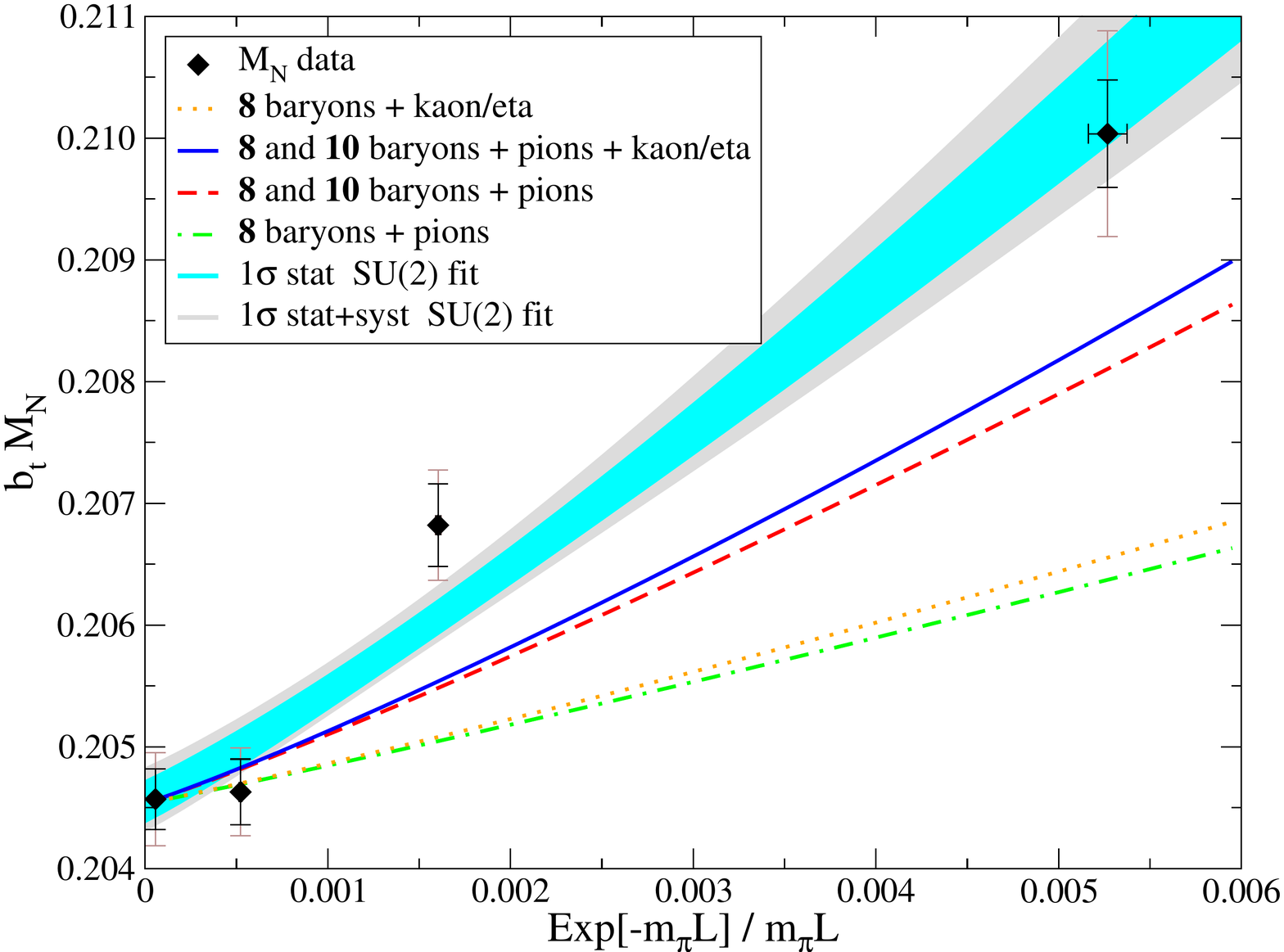}
     \caption{ The mass of the nucleon as a function of $e^{-m_\pi
         L}/( m_\pi L)$
         calculated on anisotropic gauge-field configurations 
        at a pion mass of $m_\pi\sim 390~{\rm MeV}$. 
         }
  \label{fig:NucVolPlotsu3}
\end{figure}
For  baryons, the situation is more complex because of the poor convergence properties of baryon-$\chi$PT.
Figure~\ref{fig:NucVolPlotsu3} shows the volume dependence of the nucleon mass obtained 
at a pion mass of $m_\pi\sim 390~{\rm MeV}$~\cite{Beane:2011pc}.
Its volume dependence is expected to be linear in $e^{-m_\pi L}/L$ at 
leading order in the chiral expansion, consistent with what is observed in Figure~\ref{fig:NucVolPlotsu3}.
As a function of the pion mass, the FV effects in the nucleon mass  are expected to scale as $m_\pi^3$ for fixed $m_\pi L$
(for p-regime calculations), 
and are thus expected to be significantly smaller at the physical light-quark masses~\cite{Beane:2011pc}
for a given $m_\pi L$.
However, $\chi$PT does not appear to describe the quark-mass dependence of the nucleon mass over the range 
$140~{\rm MeV} <  m_\pi <  400~{\rm MeV}$~\cite{Walker-Loud:2013yua}, 
as discussed in Section~\ref{sec:chiral}, 
so it is an open question as to whether it continues to describe volume effects at lighter masses.

It is important to quantify the FV effects and to be able to remove them from 
LQCD calculations in order to compare the spectrum of the hadrons with those of nature.
Further,  
it is vital to understand them in order  to investigate two-hadron scattering amplitudes and 
nuclear binding energies.
As nuclear binding energies are typically in the MeV range, the nucleon mass must be calculated with precision and accuracy that 
is $ < 0.1\%$.
The results shown in  Figure~\ref{fig:NucVolPlotsu3} provide guidelines for future calculations to fulfill such constraints.

While the discussions we have presented so far have been based upon the use of periodic BCs in the spatial directions,
one is  free to choose different BCs.
Periodic BCs
constrain the quark momentum  to satisfy
$\mathbf{p}=\frac{2\pi}{L}\mathbf{n}$ with $\mathbf{n}$ being an integer triplet,
and are a subset of a larger class of BCs called twisted BCs (TBCs).
 TBCs are those for which the quark fields  acquire phases $\theta_i$ at the boundaries, 
$\psi(\mathbf{x}+\mathbf{n}{\rm{L}})=e^{i{\rm{\theta}} \cdot \mathbf{n}}\psi(\mathbf{x})$,
where $0<\theta_i<2\pi$ is the twist angle in the $i^{\rm th}$ Cartesian direction~\cite{Bedaque:2004kc}.
An arbitrary total spatial 
momentum can be selected for the hadronic system by a judicious choice
of the twist angles of its valence quarks, 
$\mathbf{p}=\frac{2\pi}{L}\mathbf{n}+\frac{{\bm \phi}}{L}$,
where ${\bm \phi}$ is the sum of the twists of the valence quarks, again with 
$0<{\bm \phi}_i<2\pi$. 
TBCs have been  shown to 
be useful in LQCD calculations of 
 the low-momentum transfer behavior of form factors required in determining 
hadron radii and moments, and for calculations of the vacuum polarization contributions to the muon $g-2$,
alleviating the need for 
large-volume lattices, e.g. Ref.~\cite{Aubin:2013daa}.~\footnote{
We note that twisting is usually applied  to the valence quarks only,  
leading to a violation of unitarity in such calculations~\cite{Flynn:2005qn}. 
Nevertheless, the low energy properties of the resulting  partially quenched theory are assumed to be described by 
partially quenched $\chi$PT~\cite{Sharpe:2000bc,Sharpe:1999kj},
which can subsequently be used to correct for these effects in many cases.
}
It was recently noted that twisting can be used to reduce the FV modifications to the hadron masses~\cite{Briceno:2013hya}.
This can be accomplished by averaging the results of periodic BC and anti-periodic BC calculations, by 
twist-averaging, or by working with i-PBCs corresponding to a single twist of $\phi = (\pi/2 ,\pi/2 ,\pi/2 )$~\cite{Briceno:2013hya}.

\begin{figure}[!ht]
  \centering
  \includegraphics[width=0.5\columnwidth]{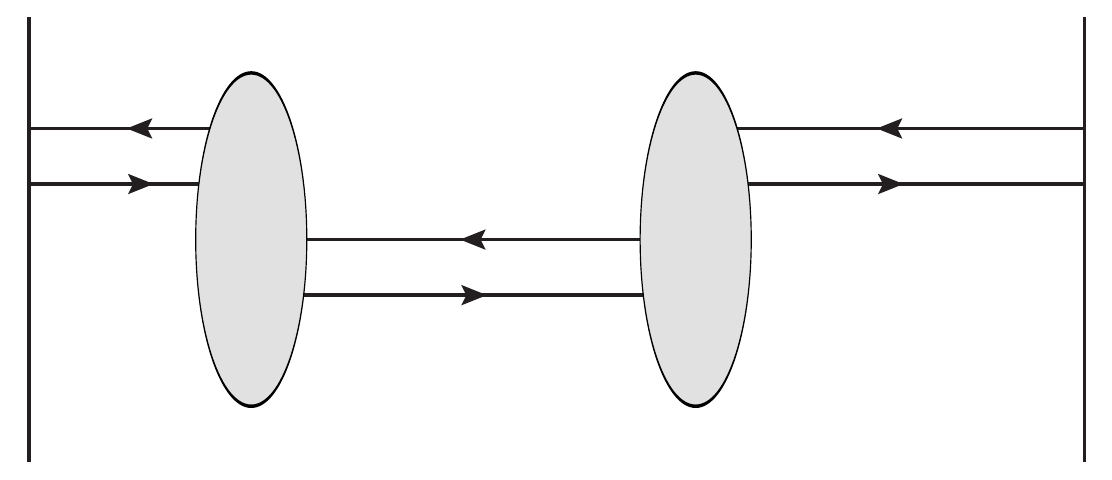}
  \caption{Thermal contributions to $\pi\pi$
    correlation functions.  The vertical lines indicate 
    anti-periodic temporal boundaries  and the
    grey regions represent the $\pi^+\pi^+$ source and sink. The solid
    lines correspond to valence quark propagators.  }
  \label{fig:thermalpipidiagram}
\end{figure}

Realistic calculations  are performed in volumes with a finite time direction, 
with the quark fields satisfying anti-periodic BCs,
corresponding to calculations of the system at a typically low, but non-zero temperature.
One effect of the finite temperature is to introduce contributions 
to the correlation function
from subsets of hadronic degrees of freedom propagating backwards in time.
These give rise to energies in the correlation function that are lower than that of the
ground state.
However, such effects are exponentially suppressed by the length of the time direction.
For interpolating functions ${\cal O}_{A,B}$, a correlation function
calculated with anti-periodic BCs on the quark-fields becomes
\begin{eqnarray}
  C_{\cal O}(t;T) & = & {1\over Z} {\rm Tr}\left[ e^{-\hat H T}\ \hat {\cal O}_A^\dagger (t) \  \hat
    {\cal O}_B(0) \right] \nonumber \\
  &=& 
  {1\over Z} \sum_{j,k} e^{-E_j T}\ e^{(E_j-E_k)t}\ \langle j|\  \hat {\cal
    O}_A^\dagger(0)\  |k\rangle  \langle k|\  \hat {\cal
    O}_B(0)\  |j\rangle
  \ \ \ ,
  \label{eq:thermal}
\end{eqnarray}
where $T$ is the length of the time-direction and $Z={\rm Tr}\left[
  e^{-\hat H T}\right]$ is the partition function.
As an example, consider an interpolating operator that
couples to the $\pi^+\pi^+$-state, which can be written in terms
of hadronic field operators as $ \hat {\cal O}(0)\ = Z_{\pi^+\pi^+}\
\pi^+\pi^+\ +\ Z_{\pi^+\pi^+\pi^0\pi^0}\ {\pi^+\pi^+\pi^0\pi^0}\
+...$, where the ellipses denote all other possible 
operators with the same quantum numbers and the 
$Z$'s are {\it a priori} 
unknown overlap factors.  
In Eq.~(\ref{eq:thermal}), this operator thus gives
non-zero values of $\langle \pi^-\pi^-|\ \hat {\cal O}(0)\
|0\rangle$, $\langle \pi^-|\ \hat {\cal O}(0)\ |\pi^+\rangle$,
$\langle 0|\ \hat {\cal O}(0)\ |\pi^+\pi^+\rangle$, 
and for all other
states with the same quantum numbers as the $\pi^+\pi^+$
source. 
Consequently, the  correlation function contains
exponentials $e^{-E\ t}$ with energies 
$E=E_{\pi^+\pi^+}$,
$E_{\pi^+} -E_{\pi^+}=0$, 
$-E_{\pi^+\pi^+}$,
$E_{\pi^+\pi^+\pi^0\pi^0}$, 
$-E_{\pi^+\pi^+\pi^0\pi^0}$,
$E_{\pi^+\pi^+K^+K^-}$,\ldots. 
In the zero temperature limit, only
those exponentials with $E\ge E_{\pi^+\pi^+}$ survive. 
States with
energies less than $E_{\pi^+\pi^+}$ 
can be interpreted as thermal excitations, for
instance arising from the process shown in
Figure~\ref{fig:thermalpipidiagram}.

As they dominate  
correlation functions at  times  $t\sim T/2$, 
 the thermal states are not simply a
curiosity that can be safely ignored.  
The amplitudes of these
states are exponentially suppressed by the temporal extent of the
lattice  times the energy of the backward going hadronic state.
Consequently, the most important thermal states involve backward
propagating pions, and
the product $m_\pi T$ must be large in order to suppress these states. 
As the chiral limit is approached, 
and the pion becomes lighter,
this  
requires increasingly large computational resources.

\subsubsection{Finite-Volume Effects and Boundary Conditions (II): Multiple Hadrons.}

Extracting hadronic interactions from LQCD calculations is significantly
more complicated than  determining the spectrum of stable particles.
This is encapsulated in the
Maiani-Testa theorem~\cite{Maiani:1990ca}, which states that $S$-matrix elements
cannot be extracted from infinite-volume Euclidean-space Green functions except
at kinematic thresholds.
This is clearly problematic from the nuclear physics perspective, as a main
motivation for pursuing LQCD is to be able to compute reactions
involving multiple nucleons.
However, Euclidean-space correlation functions  calculated in a  finite volume 
can be used to extract
$S$-matrix elements, as has been known for decades in the
context of non-relativistic quantum mechanics~\cite{Huang:1957im} and extended
to 
quantum field theory
by L\"uscher~\cite{Luscher:1986pf,Luscher:1990ux}.
The allowed energies of two particles in a  finite volume depend
in a calculable way upon their elastic scattering amplitude
for energies below the inelastic threshold.

Since  L\"uscher's original analysis~\cite{Luscher:1986pf,Luscher:1990ux}, 
there have been many works that have generalized the 
analysis to cases such as boosted systems and
those performed   with twisted BCs, 
and have made explicit the formalism for higher angular momentum channels.
For a bound system, a single hadron or a nucleus, that is compact compared with the lattice volume, 
the impact of the boundary is exponentially suppressed by the energy gap to the next lightest hadronic state.
In the large volume limit, with the impact of the lattice spacing 
parametrically diminished in the continuum limit, a bound system can be classified by its SO(3) (and other) quantum numbers.
In contrast, the (low-lying) continuum (scattering) states are intrinsically linked to the BCs, and are classified by irreps of the cubic group 
rather than SO(3).~\footnote{
One understands the recovery of SO(3) in the large-volume limit
by considering the systems at fixed energy, with finite energy resolution and high excitation in the lattice volume,
where the large multiplicity of a given cubic irrep allows for  better resolution of angular momentum~\cite{Luu:2011ep}.
} 
In volumes that are large compared with the range of the interaction, 
the energies of scattering states have a power-law dependence on the spatial extent of the lattice, with exponential corrections suppressed by the inverse of the 
range of the interaction.
For two scalar particles in the $A_1^+$ irrep of the cubic group, with  energy below the inelastic threshold, 
a direct relation between the FV energy shift, $\delta E$,
and the s-wave phase shift, $\delta_0$, 
only exists when the $l=4,6,...$ phase shifts are neglected.
In that case, the L\"uscher relation becomes
\begin{eqnarray}
q \cot\delta_0 & = & 
{2\over\sqrt{\pi} L}\ {\cal Z}_{0,0}(1;\tilde q^2)
\ \ {\rm with} \ \ 
{\cal Z}_{l,m}(s;\tilde q^2)\ =\ \sum_{\bf n} {  |{\bf n}|^l\ Y_{lm}(\Omega_{\bf n}) \over \left[ |{\bf n}|^2 - \tilde q^2 \right]^s}
\ \ \ ,
\label{eq:luscherSwave}
\end{eqnarray}
where $\tilde q = \left({L\over 2\pi}\right) q$ and  $q$ is the magnitude of the relative three momentum, derived from the energy shift 
$\delta E/2 = \sqrt{q^2+m^2} - m$ (for identical mass hadrons).
In general, for two-hadron systems, the FV energy shift receives contributions from all partial waves, and 
truncations must be made in order to extract phase-shift information. 
These truncations can be checked for self consistency by further  calculations in different volumes, or with different BCs.
However,  the extraction of phase shifts using L\"uscher's method necessarily introduces systematic uncertainties 
that require quantification.

The extention of  L\"uscher's method to multiple, coupled two-hadron channels has recently been detailed, 
e.g.~\cite{Bernard:2010fp,Doring:2012eu,Torres:2013boa,Briceno:2013bda,Briceno:2013lba,Briceno:2014oea}.
This extension has impact beyond low-energy nuclear physics, as it is essential for LQCD calculations 
in support of 
the GlueX experimental program at the Thomas Jefferson Laboratory that is focused on identifying 
gluonic excitations of hadrons and other exotic states (for recent work, see Ref.~\cite{Dudek:2013yja}).
In such coupled systems, the energy eigenvalues obtained from LQCD calculations depend upon all of the phase shifts and mixing parameters describing the 
$S$-matrix in non-trivial ways.
This is true even after neglecting the channels that do not contribute in the infinite-volume limit.
Each eigenvalue provides a  combination of scattering parameters evaluated at that energy eigenvalue.  
Generally, assumptions have to be made as to the analytic structure of the amplitude near the energy eigenvalues
in order to extract the scattering parameters. 
In some cases, this can be done in a constrained way by
appealing to EFTs to provide an approximate form of the momentum dependence~\cite{Beane:2010em,Lang:2014yfa}.
The validity and robustness of the assumed form of the scattering amplitudes has to be systematically verified, and used to 
estimate the associated systematic uncertainty.
Recently, L\"uscher's energy quantization conditions (QCs) have 
been extended to include the EM  interactions between charged hadrons~\cite{Beane:2014qha} in anticipation of 
future LQCD calculations.

\begin{figure}[ht!]
\begin{center}  
\includegraphics[scale=0.2]{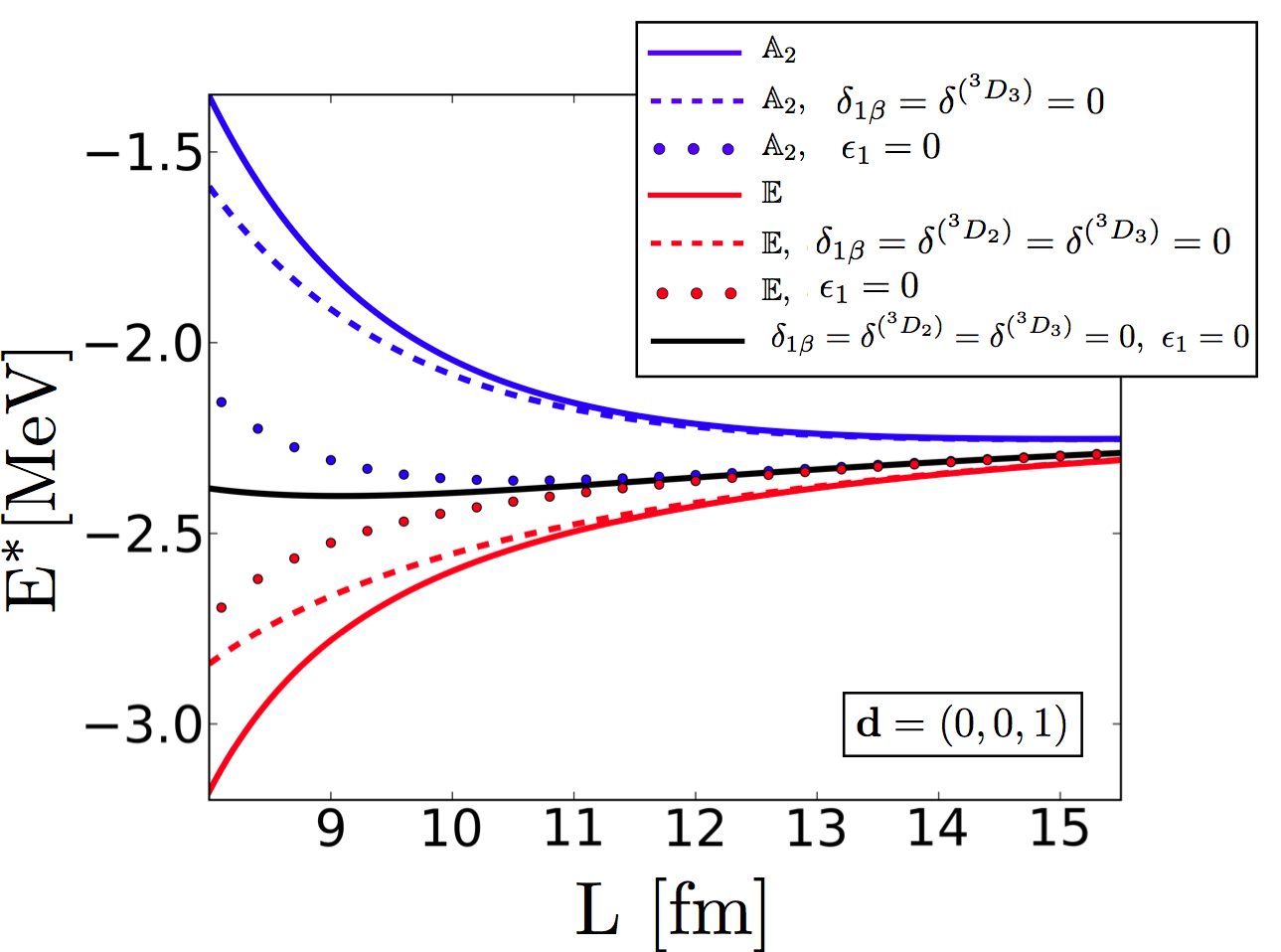}
\caption{The expected energy of two nucleons in the positive-parity
isoscalar channel (which contains the deuteron)
with boost vector
$\mathbf{d}=(0,0,1)$ as a function of
  ${\rm L}$, 
extracted from the $\mathbb{A}_2$ (red) and $\mathbb{E}$ (blue) QCs given in 
Ref.~\protect\cite{Briceno:2013bda}. 
}
\label{deut_tet}
\end{center}
\end{figure}
As the NN phase shifts and mixing parameters have been measured experimentally to relatively high precision,
studies have been performed~\cite{Briceno:2013lba} to explore the impact of truncating the  QCs
that follow from a generalization of
 L\"uscher's work,
 building upon earlier studies~\cite{Beane:2010em}.
 In the case of the deuteron, 
 the volume dependence of 
 the two lowest energy eigenvalues obtained
 from the experimentally determined phase shifts and mixing parameter
  for the system boosted with one 
 unit of lattice momentum
 (corresponding to the one-dimensional $\mathbb{A}_2$ and 
 two-dimensional $\mathbb{E}$ irreps of the cubic group) 
 are shown in Figure~\ref{deut_tet}.
 Energies  with and without the contributions from the sd-mixing parameter, $\epsilon_1$, 
 and the d-wave phase shifts in the $J=1,2,3$ channels 
 (these can all contribute because of the absence of rotational symmetry) in the vicinity of the deuteron pole,
are shown  (contributions from the higher partial waves have been neglected).
The differences in energies provide an estimate of  the impact of truncating L\"uscher's QC. 
Interestingly,  
the observed sensitivity to $\epsilon_1$ 
suggests that calculating the energy splitting between these two irreps will allow for its determination.

Beyond two-hadron systems, there are ongoing efforts aiming
to formulate relations between $S$-matrix elements describing three-body systems and the FV 
energies of three-body states~\cite{Beane:2007qr,Briceno:2012rv,Hansen:2014eka,Detmold:2008gh,Hansen:2012tf}.
However, at this point in time, only a few systems involving more than two unbound hadrons 
have been explored with LQCD~\cite{Detmold:2008yn,Detmold:2011kw,Beane:2007es,Shi:2012ay}.

\subsubsection{Chiral Extrapolations.}
\label{sec:chiral}
LQCD has revolutionized our understanding of the quark-mass dependence of hadronic observables.
It has provided precise determinations of low-energy constants defining    $\chi$PT,
and has made explicit its limitations.  
Many mesonic observables are now being calculated at, or near, the physical pion mass, although often still in the isospin limit.
Currently, the spatial volumes and temporal extents 
employed in such calculations remain somewhat small, but this situation is improving.
Only a few quantities of interest to nuclear physics have so far been calculated near the physical point, 
specifically the ground-state baryon masses, e.g. Ref.~\cite{Borsanyi:2014jba}, 
and  nucleon matrix elements of quark bilinear operators, e.g. Ref.~\cite{Green:2014xba}.
Currently, higher precision is required in all of these calculations in order to impact the experimental program.

For the most part, 
chiral extrapolations are currently still necessary and
introduce further systematic uncertainties in the predictions of LQCD calculations.
One arises from the fact that a set of coefficients have to be fit to the lattice results to perform an extrapolation at any given order in the 
chiral expansion.  
This truncation of the chiral expansion means that the $n^{\rm th}$-order  fit  potentially  deviates from QCD by an amount characterized 
by the small expansion parameter raised to the $n+1^{\rm th}$ power.

An observable that shows major deviations from  expectations of $\chi$PT  is the mass of the nucleon.
\begin{figure}[!ht]
  \centering
     \includegraphics[width=0.6\textwidth]{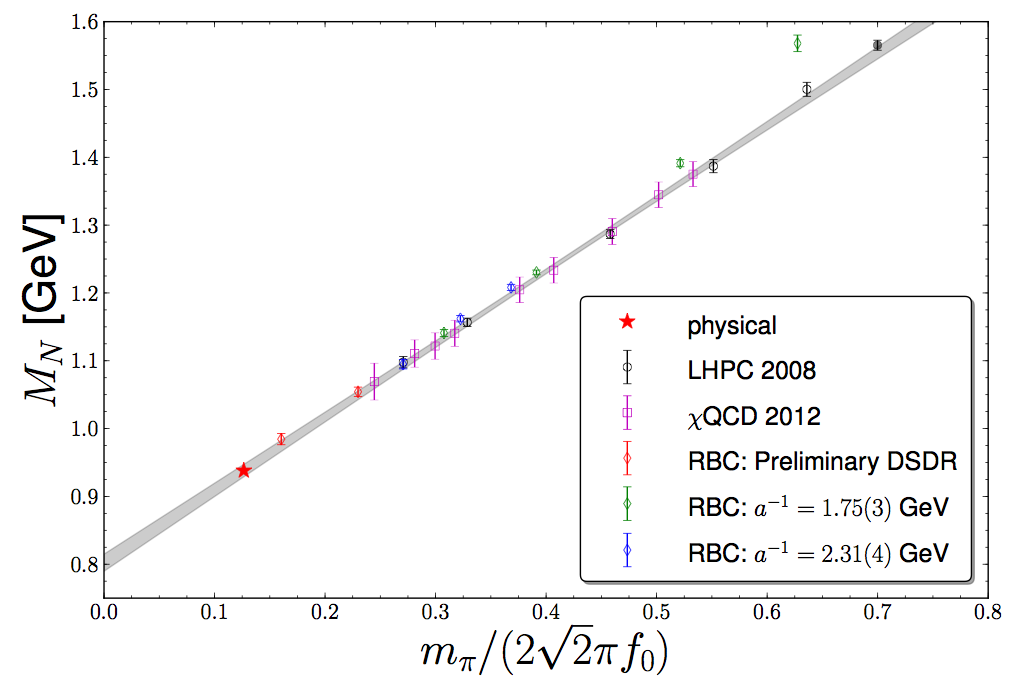}
     \caption{A compilation of LQCD calculations of the nucleon mass, as a function of $m_\pi$~\protect\cite{Walker-Loud:2013yua}. 
          [Figure reproduced with the permission of Andre Walker-Loud.]
         }
  \label{fig:MNawl}
\end{figure}
Naively, the nucleon mass has an expansion in powers of the 
light-quark masses and non-analytic contributions from loop diagrams.
Figure~\ref{fig:MNawl} shows  a compilation of the nucleon mass from LQCD calculations~\cite{Walker-Loud:2013yua}, 
which are well reproduced by a chiral dependence that is linear in the pion mass, $M_N = 800 + m_\pi~{\rm MeV}$, in conflict with expectations from $\chi$PT,
$M_N=M_0 + \alpha_1 m_\pi^2 + \alpha_2 m_\pi^3+...$.

It is now generally understood
 that, while the chiral expansion in the pseudo-Goldstone boson sector appears to converge relatively well,
 there is little indication that it is reliable for 
making predictions for the mass dependence of nucleon observables, e.g. Refs.~\cite{Beane:2004ks,Young:2005tr},
except in the  close vicinity of the chiral limit.
Therefore, there is uncertainty associated with LQCD calculations that require chiral extrapolation that is challenging to 
reliably assess. 
Because of this behavior, there have been a number of efforts to partially
resum higher orders in the chiral expansion, such as Finite-Range Regularization~\cite{Young:2005tr}.
Typically this involves identifying  a function that has the correct behavior near the chiral limit and the heavy-quark limit, 
and is a smooth function at intermediate pion masses, with parameter(s) associated with this behavior that are fit to LQCD results.
While the functional forms are not obtained from the symmetries of QCD for all quark masses, 
they do, in many instances, provide fits that agree 
well with the LQCD results, and provide predictions at the physical point with small statistical uncertainties.
The  systematic uncertainties associated with such forms are difficult to assess.
However, given  the current level of statistical precision of LQCD calculations,
they are likely  estimated sufficiently well at present.

For multi-nucleon systems, the situation is somewhat different and far less certain.  
At heavy pion masses, a pionless EFT, with only contact operators and derivatives thereof, can be used to 
describe the results of LQCD calculations and then used to make predictions for other systems 
in the periodic table~\cite{Barnea:2013uqa}.  
However, while this allows for an extrapolation in atomic number, it cannot be used for chiral extrapolations as the light-quark mass 
dependence is ``hidden'' in the coefficients of the operators.
At lighter pion masses, the chiral interactions can be used to extrapolate~\cite{Beane:2002vs,Epelbaum:2002gb,Beane:2002xf,Epelbaum:2012iu}, 
and, in fact, isolating the 
light-quark mass dependence of the nuclear forces is essential for their refinement at the physical point. 
Only one such calculation exists at present, and this is for hyperon-nucleon scattering~\cite{Beane:2012ey}.
Lattice results obtained at a pion mass of $m_\pi\sim 390~{\rm MeV}$ were used to constrain 
the hyperon-nucleon interaction at leading order in the low-energy EFT, from which predictions were made at the physical pion mass which agreed, 
within uncertainties, with phenomenological parameterizations of experimental measurements.
The uncertainties associated with this extrapolation were estimated from the size of contributions at  next-to-leading order  in the EFT expansion.

\subsubsection{Electromagnetism}

In most of the observables of interest in particle physics, EM corrections are fine-structure effects,
entering at the sub $1\%$-level.
However, 
in the structure of moderate and large nuclei,  EM becomes a leading effect.
This is because of its long-range nature, 
compared to the short range nature of the nuclear forces.
In many instances, LQCD calculations of the properties of the lowest-lying mesons
are becoming sufficiently precise that isospin violation, 
from the up- and down-quark  mass difference and the effects of EM, must be
included.  
In a few cases, hadronic properties are now being calculated
with $n_f=1+1+1+1$ 
non-degenerate quark flavors
and with fully-dynamical QED, e.g. Ref.~\cite{Borsanyi:2014jba}.  
However, most of the LQCD calculations that include EM~\cite{Blum:2007cy,Basak:2008na,Blum:2010ym,Aoki:2012st,deDivitiis:2013xla}
 do so for the valence quarks, but not in the generation of the gauge-fields,
leading to a systematic uncertainty in those results.
Including EM, while straightforward in principle, 
complicates the quark-mass  tuning  and  introduces power-law finite-volume effects.  
One advantageous feature of LQCD calculations is that 
the EM coupling constant can be chosen to be an arbitrary value, within reason,
to magnify the EM effects for the purpose of extracting them with improved precision and refining estimates of uncertainties.

\section{Hadronic Structure: Three-Point Correlation Functions}
\label{sec:3pt}

\begin{figure}[!ht]
  \centering
     \includegraphics[width=0.7\textwidth]{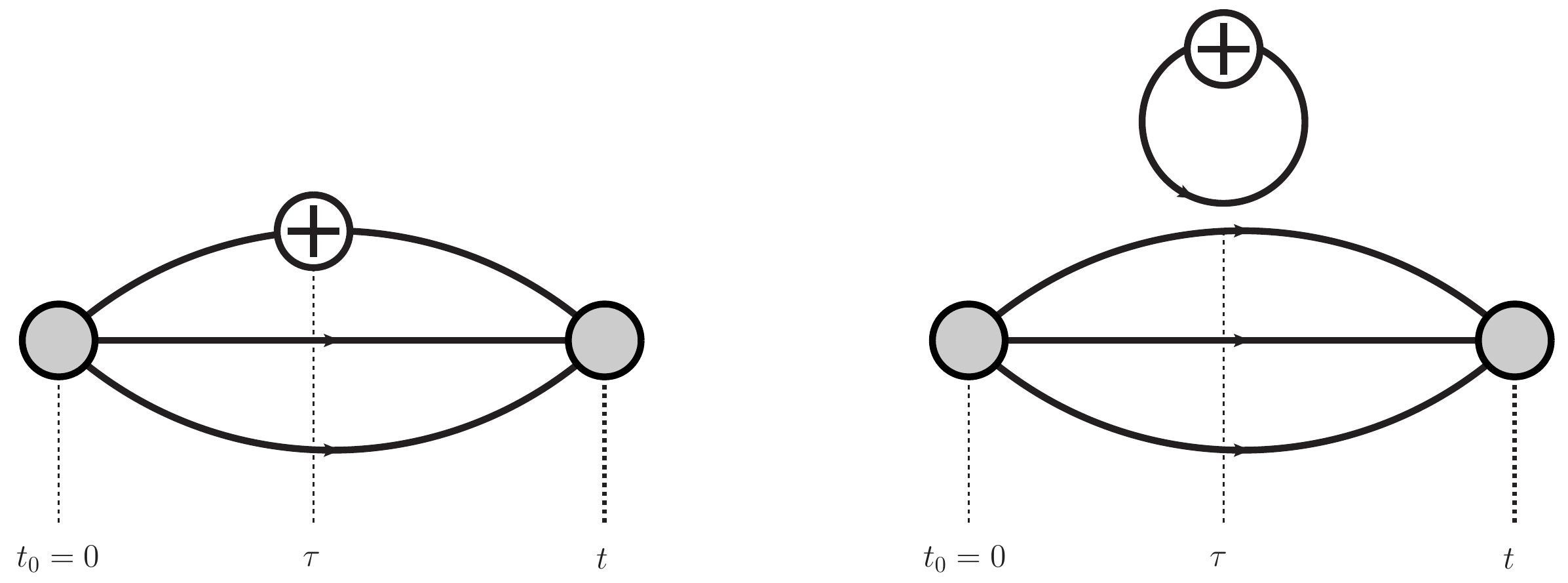}
     \caption{Contributions to three point correlation functions: connected (left) and disconnected (right).
     The shaded circles correspond to source and sink interpolating operators while the crossed circle
       represents the current insertion.  }
  \label{fig:3pt}
\end{figure}

LQCD has a much broader scope than purely the spectroscopic information available from the two-point correlators 
discussed up to now. 
A second class of observables 
of importance in nuclear physics
that have undergone extensive study are aspects of hadron structure such as 
electromagnetic form factors, moments of parton distributions and generalized parton distributions and transverse-momentum dependent
parton distributions (see Ref.~\cite{Hagler:2009ni} for a recent comprehensive review). 
These quantities correspond to matrix elements of local (and non-local) operators, 
and require the calculation of three-point correlation functions, as shown schematically in Figure~\ref{fig:3pt}. 
Using the nucleon electromagnetic form-factors as examples,  consider
\begin{eqnarray}
C_\mu({\bf p},{\bf q}; \tau,t) 
& = & 
 \sum_{{\bf z} ,{\bf y}} e^{i ( {\bf p}\cdot{\bf z}+ {\bf q}\cdot{\bf y})
}\langle 0 | \overline\chi({\bf x_0},0) J_\mu({\bf y},\tau) \chi({\bf z},t)|0\rangle
\ \ ,
\end{eqnarray}
where $\chi$ is an interpolating operator with the quantum numbers of the nucleon.
In the limit of large Euclidean time separations between the source, current insertion and sink, the insertion of complete sets 
of states on either side of $J_\mu$ shows that  this correlation function 
is given by the nucleon matrix element of this current, up to overlap and 
kinematic factors that can be extracted from simpler two point correlators,
\begin{eqnarray}
C_\mu({\bf p},{\bf q}; \tau,t) &\stackrel{0\ll \tau \ll t}{\longrightarrow}&
\langle 0 | \overline\chi(0) | N({\bf p}\rangle  \langle N({\bf p+q})|  \chi(0)|0\rangle  e^{-E_{\bf p}\tau}e^{-E_{\bf p+q}(t-\tau)} \\
&&\hspace*{1cm}\times \langle N({\bf p})| J_\mu | N({\bf p+q})\rangle
\ \ \ ,
 \nonumber
\end{eqnarray}
where the last term is the desired matrix element of the electromagnetic current between ground-state nucleons of momentum ${\bf p}$ and ${\bf p+q}$.
An additional uncertainty that is introduced in these more complicated calculations is the dependence of the results on the 
source--operator and operator--sink time separations rather than just on the source--sink separation. 
As with two-point functions, the source couples to all eigenstates of the specified quantum numbers, 
but the high lying states are exponentially damped out in Euclidean time. 
However, the inserted current  can couple back to the excited states, introducing more contamination. It is only in the limit of both $\tau$ and $t-\tau$ 
being large that the matrix elements can be extracted simply. 
Unfortunately this requires temporal separations of the source 
and sink that are larger than those needed for two-point correlation functions  
and so calculations of 
matrix elements  are expected to be 
both noisier than, and  subject to more excited state contamination than,
 the corresponding two-point functions. A number of techniques have been explored to 
address this issue, 
for instance,
through the use of matrices of correlators and multiple state fits
(see, for example, Ref.~\cite{Green:2014xba,Bhattacharya:2013ehc,Jager:2013kha}).

In many studies, uncertainties were introduced into  calculations of three-point correlation functions
by the omission of  quark-line disconnected diagrams (the right-hand diagram in Figure~\ref{fig:3pt}). 
In these terms, the quark field creation and annihilation operators in the inserted current self contract, 
resulting in a propagator from the insertion point to itself, and are present whenever the current under consideration has a flavor-singlet component. 
All-to-all propagator techniques,  discussed previously, are used to calculate such contributions, and require substantial computational resources. 
In the single nucleon sector, sophisticated all-to-all propagator techniques have recently allowed complete  calculations of the  proton
electromagnetic form factors, e.g. Ref.~\cite{Meinel:Lattice:2014},
and of the spin decomposition of the nucleon, e.g. Ref.~\cite{Deka:2013zha}.
Ongoing work to apply these methods to nuclei is underway.

\section{Error Budgets}
\begin{figure}[!ht]
  \centering
     \includegraphics[width=0.5\textwidth]{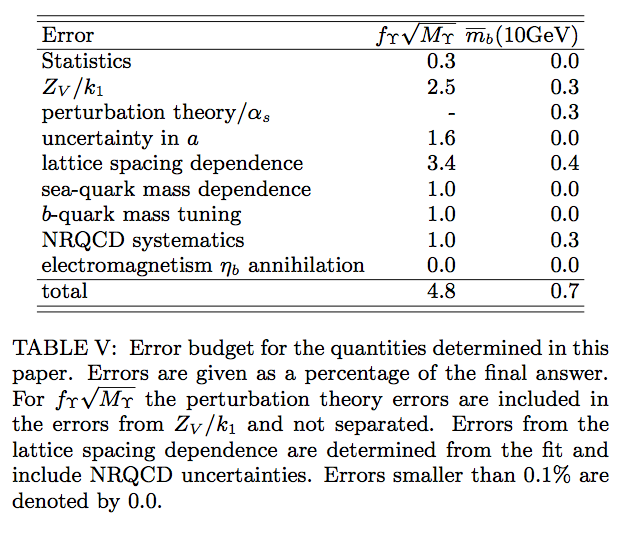}
     \caption{The error budget for the $\Upsilon$ decay constant and b-quark mass 
     determined in the calculations of Ref.~\protect\cite{Colquhoun:2014ica}.
          [Table reproduced with the permission of HPQCD collaboration.]
     }
  \label{fig:EBL}
\end{figure}

The results of a series of LQCD calculations of a given observable are presented as a central value, 
a statistical uncertainty and a systematic uncertainty 
(for correlated quantities, this is generalized to a central point and associated region in a multi-dimensional space).
However, a great deal of information is contained in the various contributions to both such uncertainties.  
In  precision calculations of weak matrix elements, and other fundamental quantities in particle physics, 
it is now standard to  provide an ``error budget'' by  tabulating all of the sources of uncertainty,  
an example of which is shown in Figure~\ref{fig:EBL}~\cite{Colquhoun:2014ica}.
Included in the contributions to the total uncertainties are those from the statistics of the calculation of the quantity, 
from lattice perturbation theory (matching  continuum QCD to  lattice QCD),
from 
chiral, continuum and infinite-volume extrapolations,
from determining the lattice spacing, 
from tuning the quark masses, and from 
the absence of EM.
In calculations of simple quantities,
sufficient computational resources are now available to numerically determine 
and control most of
these uncertainties,
through calculations with multiple lattice volumes, spacings and quark masses.  
Even in 
cases where complete calculations such as these are prohibitively computationally expensive,
a clear and complete error budget presenting all of the uncertainties, even if they result from estimates 
based upon dimensional analysis or EFT arguments,
is  informative.

\section{Summary}

Lattice quantum chromodynamics 
is a numerical technique with which to calculate strong-interaction observables in the low-energy regime
with uncertainties that can be fully quantified and systematically reduced. 
In this article, we have attempted to summarize all of the sources of uncertainty that arise in starting from the 
handful of parameters that define QCD  
(the quark masses and the strong interaction length scale)
and using the numerical machinery of LQCD to make
predictions for  low-energy observables, such as the meson and baryon spectra, the structure of the nucleon, 
and the masses and interactions of nuclei.
At the time of writing this article,  relatively simple quantities are being calculated 
at, and near, the physical light-quark masses, essentially 
eliminating one of the major uncertainties that has been 
present in
 the field for many years - the chiral extrapolation.  
While the lattice volumes and spacings
that are computationally accessible at the present time
are not ideal for nuclear physics, with increased computational resources
and algorithmic improvements, 
we expect  that  
calculations of many quantities of importance to nuclear physics with fully quantified statistical and systematic uncertainties 
will become routine in the near future.

\end{document}